\documentclass[twocolumn]{aastex63}

\usepackage{hyperref}
\usepackage{xcolor}
\usepackage{url}
\usepackage{amssymb}
\usepackage{amsmath}
\usepackage{float}
\hypersetup{
    colorlinks,
    linkcolor={red!50!black},
    citecolor={blue!50!black},
    urlcolor={black} 
}
\usepackage{mathrsfs}
\usepackage{nicefrac}

\newcommand{\kms}{km $\rm s^{-1}$}

\graphicspath{{./}{figures/}}

\received{August 23, 2020}
\accepted{December 26, 2020}
\submitjournal{AJ}
\shorttitle{Contrast and Temperature Limits to HRCCS}
\shortauthors{Finnerty et al.}

\begin{document}

\title{Contrast and Temperature Dependence of Multi-Epoch High-Resolution Cross-Correlation Exoplanet Spectroscopy}

\correspondingauthor{Luke Finnerty}
\email{lfinnerty@astro.ucla.edu}

\author[0000-0002-1392-0768]{Luke Finnerty}
\affiliation{Division of Physics, Mathematics, and Astronomy, California Institute of Technology \\
Pasadena, CA 91125, USA}

\author[0000-0002-9943-6124]{Cam Buzard}
\affiliation{Division of Chemistry and Chemical Engineering, California Institute of Technology \\
Pasadena, CA 91125, USA}

\author{Stefan Pelletier}
\affiliation{Institute for Research on Exoplanets, Universit{\'e} de Montr{\'e}al, Montreal, QC, Canada}

\author[0000-0003-4451-2342]{Danielle Piskorz}
\affiliation{Division of Geological and Planetary Sciences, California Institute of Technology \\
Pasadena, CA 91125, USA}

\author{Alexandra C. Lockwood}
\affiliation{Space Telescope Science Institute, Baltimore, MD 21218, USA}

\author[0000-0003-4384-7220]{Chad F. Bender}
\affiliation{Department of Astronomy and Seward Observatory, University of Arizona, Tuscon, AZ 85721, USA}

\author[0000-0001-5578-1498]{Bj{\"o}rn Benneke}
\affiliation{Institute for Research on Exoplanets, Universit{\'e} de Montr{\'e}al, Montreal, QC, Canada}

\author[0000-0003-0787-1610]{Geoffrey A. Blake}
\affiliation{Division of Geological and Planetary Sciences, California Institute of Technology \\
Pasadena, CA 91125, USA}
\affiliation{Division of Chemistry and Chemical Engineering, California Institute of Technology \\
Pasadena, CA 91125, USA}

\begin{abstract}
While high-resolution cross-correlation spectroscopy (HRCCS) techniques have proven effective at characterizing the atmospheres of transiting and non-transiting hot Jupiters, the limitations of these techniques are not well understood.  We present a series of simulations of one HRCCS technique, which combines the cross-correlation functions from multiple epochs, to place temperature and contrast limits on the accessible exoplanet population for the first time.  We find that planets approximately Saturn-size and larger within $\sim$0.2 AU of a Sun-like star are likely to be detectable with current instrumentation in the $L$-band, a significant expansion compared with the previously-studied population.  Cooler ($ \rm T_{eq} \leq 1000$ K) exoplanets are more detectable than suggested by their photometric contrast alone as a result of chemical changes which increase spectroscopic contrast. The $L$-band CH$_4$ spectrum of cooler exoplanets enables robust constraints on the atmospheric C/O ratio at $\rm T_{eq} \sim 900K$, which have proven difficult to obtain for hot Jupiters.  These results suggest that the multi-epoch approach to HRCCS can detect and characterize exoplanet atmospheres throughout the inner regions of Sun-like systems with existing high-resolution spectrographs. We find that many epochs of modest signal-to-noise ($\rm S/N_{epoch} \sim 1500$) yield the clearest detections and constraints on C/O, emphasizing the need for high-precision near-infrared telluric correction with short integration times.
\end{abstract}



\section{Introduction} \label{sec:intro}
Despite the detection of more than 4000 exoplanets in the past 25 years, very little is known about most exoplanets aside from an orbital period and a radius or minimum mass. 
This is particularly frustrating for exoplanet populations such as warm and hot Jupiters, which have no analogue in our own solar system and therefore pose interesting questions for theories of planet formation and evolution. 
The formation of hot Jupiters has been an open topic since the first exoplanet was discovered around a Sun-like star \citep{mayor}. 
Recent theoretical work and analysis of multi-planet systems has raised the possibility that hot Jupiters may form in-situ \citep[e.g.][]{Batygin_2016}, contrasting with earlier work suggesting formation beyond the H$_2$O snow line followed by inward migration and tidal circularization \citep[e.g.][]{Lin1996}. 

Intermediate-temperature gaseous planets ($\rm T_{eq} \approx 1000$ K) are particularly interesting in this context, as the tidal circularization timescale for more widely-separated planets is often much longer than the age of the system \citep{correia_2010}. 
Furthermore, while nearly all known hot Jupiters (period $\rm P < 10$ days) lack close companions, half of warm Jupiters ($\rm 10 < P < 200$ days) have such companions, suggesting different evolutionary processes may be responsible for hot versus warm giant planets \citep{Huang_2016}. 
In recent years, a range of formation mechanisms have been proposed for warm Jupiters. 
Many of these theories involve planet-planet scattering interactions based on the high companion fraction and moderate eccentricities of known warm Jupiters \citep{Petrovich_2016, Masuda_2017, Anderson_2017, Anderson2020}, but it is not clear whether such interactions are responsible for delivering these planets to their observed location from beyond the snow line, or whether these planets initially form in-situ. 
It is also unclear how the hot and warm Jupiter populations are related.
    
Detailed constraints on exoplanet atmospheres have the potential to resolve the ambiguous origins of both hot and warm Jupiters. 
The ratio of carbon to oxygen (C/O ratio) in gas versus solids in a static protoplanetary disk has been shown to vary consistently with orbital distance \citep{Oberg_2011}. 
This suggests the atmospheric C/O ratio of giant exoplanets may offer insight into where and how these planets form \citep{Madhusudhan_2014}, though accretion of solid material significantly complicates the relationship \citep[e.g.][]{espinoza_2017}. 
With sufficiently accurate models of accretion histories, atmospheric metallicity and C/O ratio could allow planets which formed beyond the water snow line and migrated to their present position to be distinguished from planets which initially formed in the inner disk. 
This would help answer the question of in-situ formation versus migration mechanisms and shed light on the possible difference in formation mechanisms for hot and warm Jupiters.
Making these measurements requires the ability to simultaneously detect carbon and oxygen-bearing species in both stellar and exoplanetary atmospheres and estimate their relative abundances.

Treating the star/planet system as a spectroscopic binary offers an avenue for planet characterization that is less dependent on orbital inclination compared with transit-based techniques. 
Hot Jupiters are sufficiently bright in the thermal infrared to be detected using high-resolution spectrographs and cross-correlation with a model planet spectrum, an approach referred to as high-resolution cross-correlation spectroscopy (HRCCS).  
In the past decade, this has been used to detect molecules including H$_2$O, CO, TiO, HCN, and CH$_4$ in both transiting and non-transiting hot Jupiter atmospheres \citep[e.g.][]{Snellen2010, Rodler2012, lockwood,  Brogi2014, Birkby_2017, nugroho_2017, Hawker_2018, Guilluy_2019}, as well as provide rough constraints on the planetary C/O ratio \citep{Brogi_2017, Piskorz2018}. 
As these techniques yield a value for the target exoplanet's radial velocity semi-amplitude $\rm K_p$, these detections have the added benefit of breaking the mass/inclination degeneracy of RV-detected planets, giving the true planet mass. 
An important difference compared with other exoplanet characterization techniques is that the cross-correlation function uses all lines present in both the observed spectrum and a model spectrum to detect the planet. 
Placing constraints on the target atmosphere therefore requires assessing how the cross-correlation function varies as a result of changes in the model spectrum used to perform the correlation.  
For example, \citet{lockwood} uniquely identified the presence of H$_2$O in the atmosphere of $\tau$ Boo b by performing the cross-correlation with a set of models containing only a single molecule's spectrum, only making a detection when the H$_2$O model was used. 
\citet{Piskorz2018} extended this approach to attempt constraints on the C/O ratio, metallicity, and incident stellar flux in the atmosphere of KELT-2Ab by correlating with a series of planet models with different parameters.

Two techniques for HRCCS have been developed. 
The ``1D" approach, described in \citet{Snellen2010}, is effective for very close-in planets whose projected orbital velocity changes significantly over a few (typically 5--7) hours of observation.  
Over short timescales, stellar lines remain fixed in wavelength, while planet lines shift due to orbital motion. 
A 1D cross-correlation with a planet spectral template can then be used to identify lines that shift over the course of the observation, effectively measuring the planet's radial acceleration, which yields the planet radial velocity semi-amplitude $\rm K_p$. 
As stellar lines are fixed in wavelength over the time series, the impact of such features on the planet detection is minimal, greatly simplifying the analysis procedures.

An alternative approach, known as the ``2D" or ``multi-epoch" technique, was first described in \citet{lockwood}.  
Rather than taking a nearly continuous series of exposures over many hours, this approach takes shorter observations spread over several nights, during each of which the planet features are effectively fixed in wavelength. 
In this technique, the extreme contrast between stellar and planetary lines and fixed planetary velocity within an observation requires a 2D cross-correlation using both stellar and planetary model spectra to simultaneously identify the fixed stellar and planetary velocities.
The cross-correlation surfaces for each night are then combined to determine a best-fit planetary velocity semi-amplitude. 
As a result, the 2D technique is sensitive to the total variation in the planetary radial velocity over the planetary orbital period, rather than only the change over a few hours of observation. This should enable the detection of slower-moving planets on longer orbital periods which are inaccessible to the 1D technique, as the planetary radial acceleration is a much stronger function of the semimajor axis than the planetary radial velocity. 

Although both HRCCS techniques have proven successful, the limitations of the 2D approach are not well understood. 
While the 1D technique is limited to close-in planets by the large radial acceleration required, the total orbital velocity change measured by the 2D technique is significantly larger than the resolution of existing instruments for circular orbits up to $\sim 4$ AU from a Sun-like star. 
This suggests that the detection limit for the 2D technique will be determined by the physical and chemical properties of the system, rather than instrument resolution. 
A detailed understanding of how various physical and chemical factors, such as photometric contrast, star/planet equilibrium temperatures, and planet atmospheric composition affect detection could enable the characterization of a broad population of exoplanets which are too widely separated from their host stars for the 1D technique to be effective but too close to be detected with direct imaging. 
A key advantage over transit-based characterization techniques is the much weaker inclination requirements allow cross-correlation techniques to be used on many more targets, particularly at larger orbital separations.

In this paper, we present a series of simulations that begin to address the shortcomings in our knowledge of the limits to the 2D multi-epoch approach, based on observations taken with Keck-NIRSPEC2.0 \citep{McLean1998, martin2018}, focusing primarily on the photometric contrast, equilibrium temperature, and C/O ratio of the target exoplanet.  
We consider in detail the factors affecting exoplanet detectability in the $L$-band (2.9--3.7 $\mu$m) using the 2D multi-epoch approach from \citet{lockwood}, simulating observational, instrumental, and physical inputs as described in \citet{Buzard_2020}. 
We describe these simulations in detail in Section \ref{sec:sims}, and discuss the resulting limits on planet size/photometric contrast in Section \ref{sec:rsim} and impact of equilibrium temperature on detection in Section \ref{sec:tlim}. 
Based on the findings in Section \ref{sec:rsim}, we assess the ability of the cross-correlation technique to constrain C/O ratio in Section \ref{sec:co}, focusing on planets cooler than those previously studied. 
Instrumental factors in detection are assessed through simulations in Section \ref{sec:inst}. 
Section \ref{sec:disc} discusses the results of these simulations and the implications for future multi-epoch HRCCS observations.  
Section \ref{sec:conc} summarizes our findings.

\section{Methods}\label{sec:sims}
\begin{figure*}
    \centering
    \noindent\includegraphics[width=39pc]{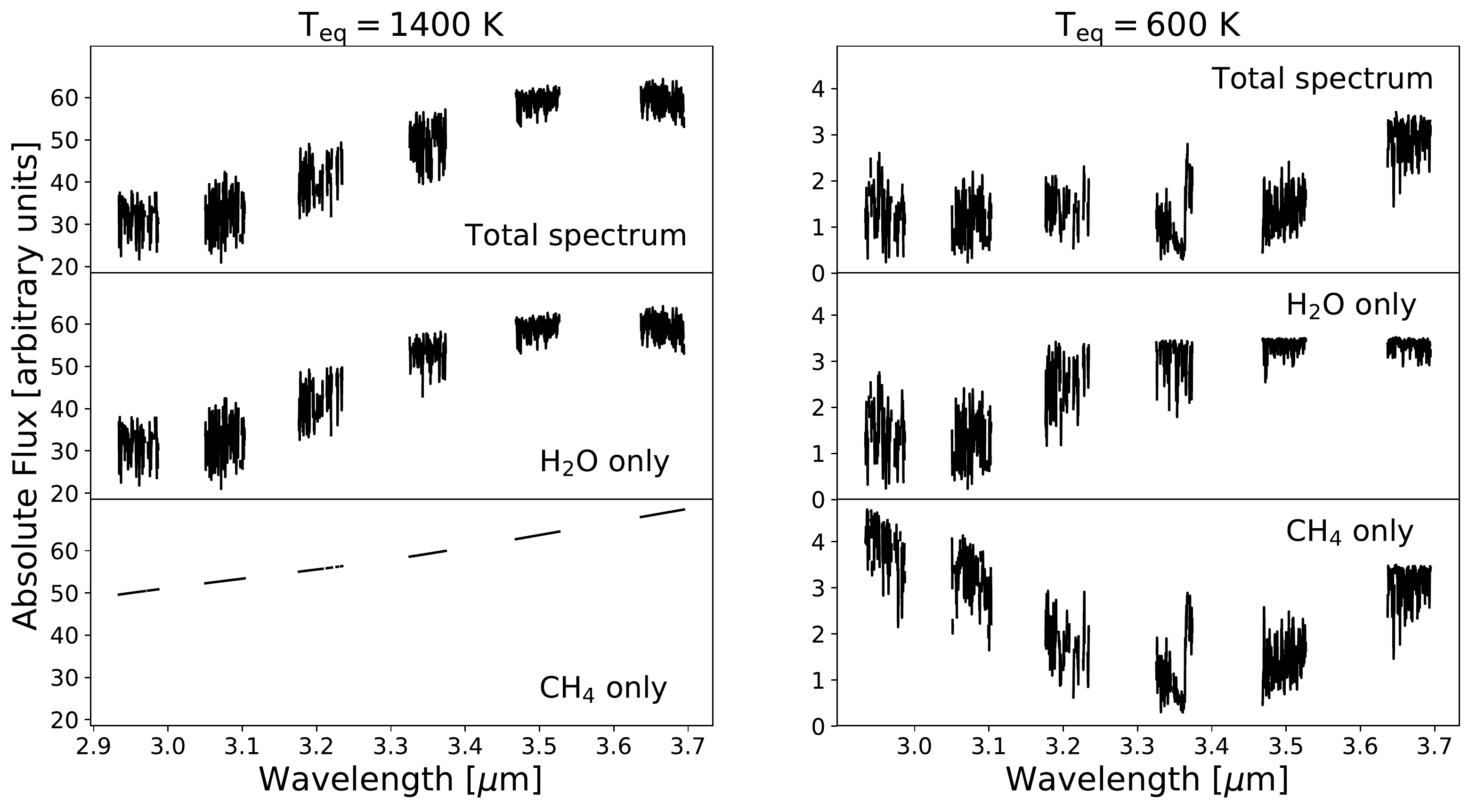}
    \caption{Model $L$-band planet spectra for $\rm T_{eq}$ of 1400 K (left) and 600 K (right), masking regions not observed with NIRSPEC. The top row plots the total spectrum including all species, while the middle and bottom rows plot the individual H$_2$O and CH$_4$ spectra respectively. While the spectrum is dominated by weak H$_2$O features at 1400 K, the 600 K spectrum is dominated by deep CH$_4$ features. The dramatic difference in spectra indicates the spectroscopic contrast evolves differently with $\rm T_{eq}$ than the photometric contrast, benefiting the cross-correlation detection of cooler planets.}
    \label{plmodspec}
\end{figure*}

\begin{figure}
    \centering
    \noindent\includegraphics[width=20pc]{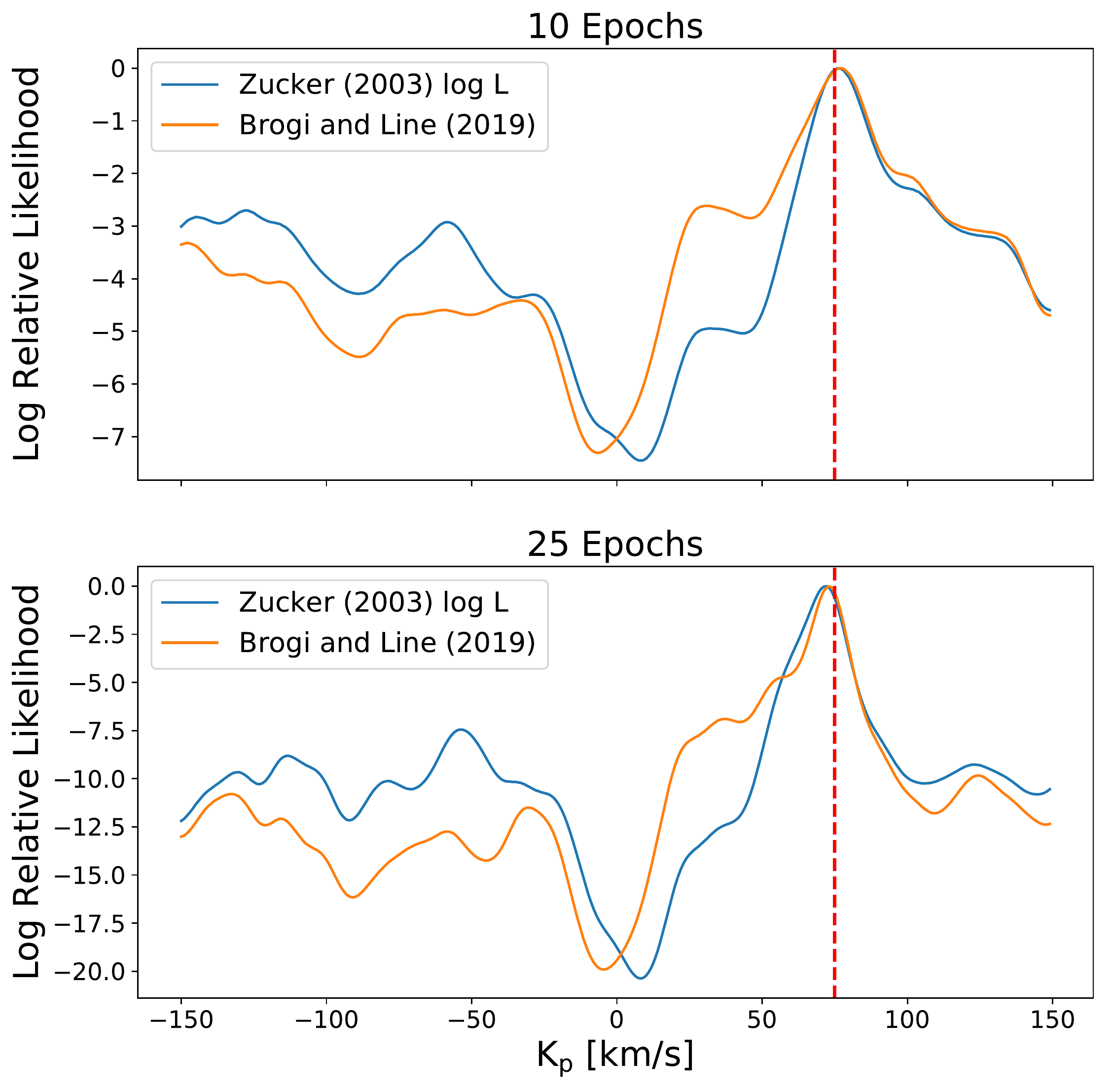}
    \caption{Comparison of the \citet{zucker2003} $\log L$ and \citet{brogi2019} techniques for combining cross-correlations with 10 and 25 epochs. Side peaks near the exoplanet velocity due to features in the stellar spectrum are stronger in the \citet{brogi2019} approach compared with the \citet{zucker2003} approach, causing us to focus on the \citet{zucker2003} method for this work.}
    \label{brogizuckercomp}
\end{figure}

High-fidelity simulations of 2D multi-epoch HRCCS were first presented in \citet{Buzard_2020}. 
The description of the simulated spectra is summarized here for completeness. 
These simulations make use of PHOENIX stellar models from \citet{phoenix}, interpolating over equilibrium temperature $\rm T_{eq}$,  metallicity $z$, and surface gravity $\log g$ in order to match the desired stellar properties. 
For all simulations, a model stellar spectrum based on the Sun-like star HD 187123 A from \citet{Buzard_2020} is used, with effective temperature $\rm T_{eff} = 5815$ K, surface gravity $\log g = 4.359$, and metallicity $\rm [Fe/H] = 0.121$. 
Planet models are created with SCARLET \citep{benneke, benneke_2019}, and can be generated with specified values for $z$, $\log g$, C/O ratio, and $\rm T_{eq}$. 
The planet models are similar to those used in \citet{Buzard_2020} for HD 187123 b, though with a noninverted atmospheric temperature-pressure profile.  
The lack of an inverted T--P profile at the temperatures simulated is supported by both theory and observations \citep[e.g.][]{Fortney_2008, Line_2016}.  
Unless otherwise noted, planet models have a solar C/O ratio and metallicity, Jupiter surface gravity, $\rm R = 1.0\ R_J$, and $\rm T_{eq} = 1400\ K$. 
Figure \ref{plmodspec} plots simulated planet spectra with equilibrium temperature of 1400 K and 600 K over the wavelength ranges simulated, which are based on prior observations with Keck/NIRSPEC2. 

In general, 10 or 25 epochs were simulated, evenly spaced in orbital phase, with a per-epoch signal-to-noise ($\rm S/N_{epoch}$) of 1500 for 25 epochs and 2500 for 10 epochs. 
The 10--epoch, high $\rm S/N_{epoch}$ case is meant to approximate a large telescope such as Keck, while the 25--epoch case approximates a smaller, less oversubscribed telescope which can obtain more epochs but with lower S/N in a given integration time. 
These simulations represent a significantly larger data set than previous observations, and are intended to set limits for the many-epoch, lower signal-to-noise observing strategy suggested in \citet{Buzard_2020}. 

To create the simulated observed spectra, stellar and planetary model spectra are first Doppler-shifted based on the systemic velocity of the target at the time of observation and the planet orbital velocity:
\begin{equation}
\begin{split}
v_{pri} & = v_{rad} - v_{bary} \\
    v_{sec} & = K_p \sin \frac{2 \pi t_{obs}}{P} + v_{rad} - v_{bary}\\
\end{split}
\end{equation}
Where $v_{pri}$ is the velocity of the star, $v_{sec}$ is the velocity of the planet, $v_{rad}$ is the systemic radial velocity, $v_{bary}$ is the velocity of the observer at the time of observation with respect to the Earth-Sun barycenter in the direction of the target system, $\rm K_p$ is the planet radial velocity semi-amplitude, and $t_{obs}$ is the time of observation measured from inferior conjunction. 
The stellar reflex velocity is much smaller than the velocity precision of NIRSPEC, and therefore not included.
Barycentric velocities for simulations were taken from a selection of values evenly spaced between -15 and 15 \kms, and were the same for all simulations of the same number of epochs. 
An orbital period of 3.1 days and $\rm K_p = 75$ \kms\ were chosen based on typical properties for a hot Jupiter system around a Sun-like star. 
Observation times were selected to give even phase sampling of the planetary orbit. 
Changes in period do not impact the final cross-correlation surface, provided $\rm K_p$, $v_{bary}$, and observed orbital phases are fixed, and these simulations should therefore apply to longer orbital periods as well.

The spectra are then scaled to the desired radii and combined, interpolating onto the wavelength grid of the planet model. 
Stellar continuum is removed with a third--order polynomial fit over 2.8--4.0 $\mu$m in wavenumber space, the same procedure as is used for the stellar spectral template in the cross-correlation routine. 
The combined model is convolved with an instrument profile determined from prior NIRSPEC2 observations and interpolated onto a NIRSPEC2 wavelength grid, at which point regions with observed telluric absorption of more than 25 percent are masked. 
Finally, Gaussian noise is added at the desired per-pixel signal-to-noise ratio of the observation.

The combined simulated spectrum is passed into a 2D cross-correlation routine to identify the planet signal and compute a log-likelihood $\log L$. 
For this work, the \citet{zucker2003} $\log L$ approach as outlined in \citet{Buzard_2020} is used to convert 2D cross-correlation functions into log-likelihoods:
\begin{equation}
    \log L = -\frac{N}{2}\log(1-R^2)
\end{equation}
Where $N$ is the number of points in the spectra and $R$ is the 2D cross-correlation function, calculated as described in \citet{zuckertodcor}. An alternative technique to combining 1D cross-correlation functions was described in \citet{brogi2019}, and adapted for 2D cross-correlation functions in \citet{Buzard_2020}:
\begin{equation}
    \log L = -\frac{N}{2} \left\{ \log(\sigma_f \sigma_g) + \log\left[\frac{\sigma_f}{\sigma_g} + \frac{\sigma_g}{\sigma_f} - 2R\right] \right\}
\end{equation}
Where $\sigma_f$ is the variance of the observed spectrum and $\sigma_g$ is the combined variance of the two correlation templates. 
We compare these two approaches in 10 and 25 simulated epochs, shown in Figure \ref{brogizuckercomp}.  
The peak corresponding to the planet is similar in both cases, but as discussed in \citet{Buzard_2020}, off-peak correlation features are stronger relative to the true peak using the \citet{brogi2019} technique, particularly the feature near 25 \kms. 
We therefore focus on the \citet{zucker2003} technique as it gives a clearer detection of the true planet peak in the $\log L(\rm K_p)$ curve.
In Section \ref{sec:rsim}, we introduce a technique to correct for these off-peak correlation features, which leads to nearly identical performance between the two approaches. We continue to prefer the \citet{zucker2003} technique since it is unlikely star-only simulations will be able to correct off-peak structure with the same precision in real observations.

In cases where there is significant non-planetary structure in the $\log L$ surface, it is useful to consider the relative likelihood surface, $\log RL$:
\begin{equation}\label{rleqn}
    \log RL = \log L - \log \Bar{L}
\end{equation}
Where $\log \bar{L}$ is the log-likelihood function arising from features unrelated to the planet spectrum, such as stellar features or telluric residuals. The subtraction in log space is equivalent to a division in linear space, and is effectively normalizing the planetary log-likelihood function based on prior knowledge of the structure of the correlation space. 

\citet{Buzard_2020} showed that simulations constructed in this way could closely replicate actual NIRSPEC observations for HD 187123, a system containing a hot Jupiter orbiting a Sun-like star. 
Notably, the guided PCA telluric removal is sufficiently complete that features included in model telluric spectrum are almost entirely removed. 
This allows us to generate simulated observations without simulating the full telluric removal procedure. 
Unlike observations, these simulations will not include off-peak structure due to differences between the observed and model telluric spectra, but such features do not dominate the log-likelihood surface.
A range of exoplanet systems can be simulated to assess the role of various observational and physical factors in planet detection through the 2D multi-epoch approach. 
Different stellar and planetary properties can be mimicked by changing the model spectra used to create the simulations, though we can only compare with observational results for hot Jupiters.
Observational properties can be modeled by changing the signal-to-noise ratio, the number of epochs, and the sampling of the orbital phase by the simulated epochs. 
Orbital properties are modeled by changing the period and amplitude of the planet radial velocity variation to match the desired inclination and semi-major axis. 

\section{Results}\label{sec:res}
\subsection{Photometric Contrast Simulations}\label{sec:rsim}
\begin{figure*}
    \centering
    \noindent\includegraphics[width=40pc]{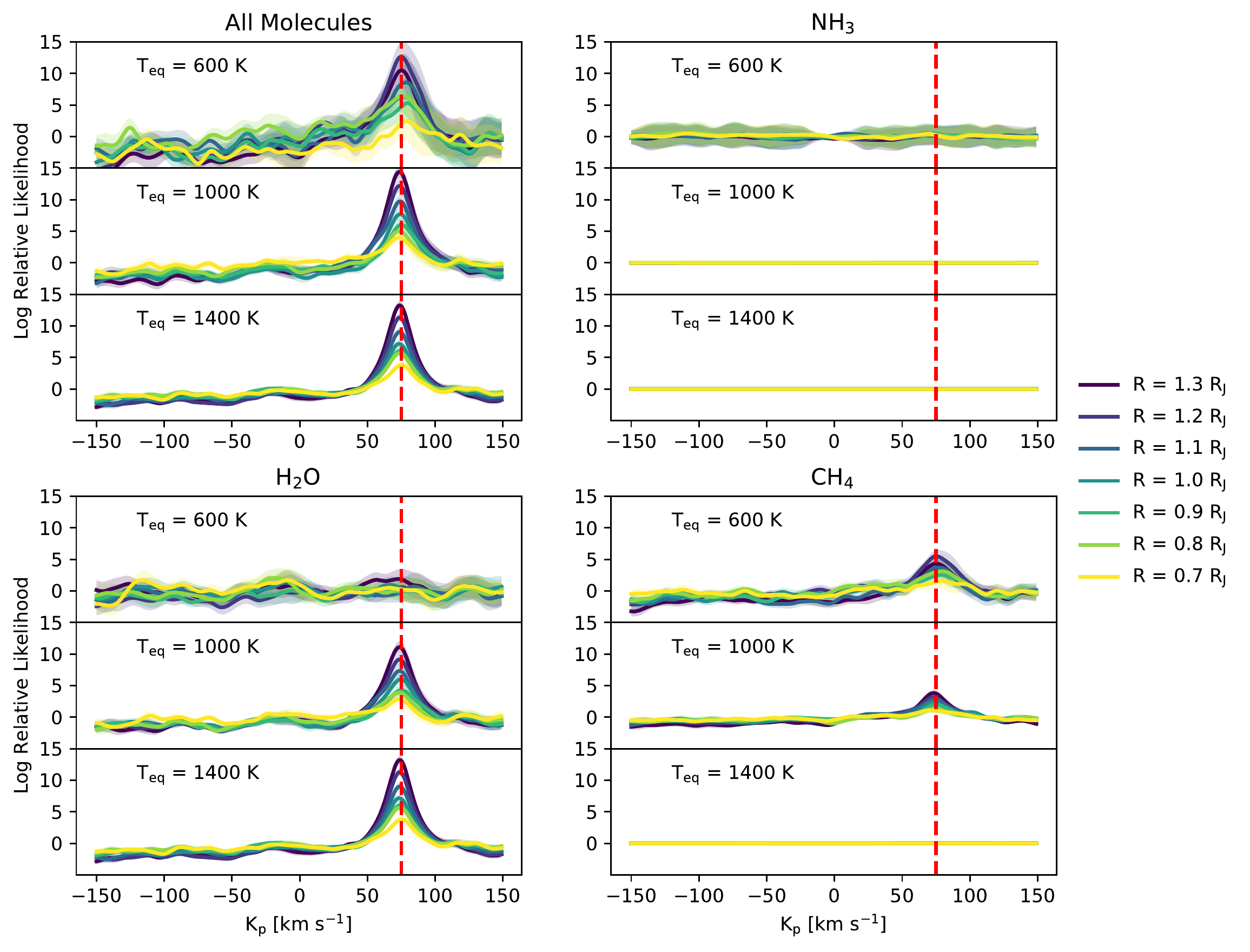}
    \caption{Relative likelihood curves for 25 observed epochs with $\rm S/N_{epoch} = 1500$ of planets with equilibrium temperatures of 600 K, 1000 K, and 1400 K and photometric contrasts corresponding to radii ranging from 0.7 $\rm R_J$ to 1.3 $\rm R_J$. Structured correlation has been removed by subtracting the log-likelihood curve of a star-only simulation, and all curves have been shifted to have an off-peak median of zero for plotting.  The top left panel plots the combined relative likelihood function of all molecules in the SCARLET planet model, while the remaining panels plot the relative likelihood curves from individual molecules.  While H$_2$O dominates the full model detection at 1400 K, CH$_4$ dominates at 600 K, transitioning around 1000 K.  The detection of cooler planets is stronger than expected based on the difference in photometric contrast compared with warmer planets.}
    \label{radgridall}
\end{figure*}

\begin{deluxetable*}{ccccccc}\centering
\tablewidth{0pt}
\tabletypesize{\scriptsize}
\tablehead{\colhead{Planet Radius} & \colhead{T = 600, 10 Epochs} & \colhead{T = 600, 25 Epochs} & \colhead{T= 600, 50 Epochs} & \colhead{T = 1400, 10 Epochs} & \colhead{T = 1400, 25 Epochs} & \colhead{T= 1400, 50 Epochs} \\ $\rm R_J$ & $\rm S/N_{epoch}$ = 2500 & $\rm S/N_{epoch}$ = 1500 & $\rm S/N_{epoch}$ = 1500 & $\rm S/N_{epoch}$ = 2500 & $\rm S/N_{epoch}$ = 1500 & $\rm S/N_{epoch}$ = 1500}
\startdata
1.3 & 4.6 & 9.7 & 19.7 & 5.4 & 12.3 & 24.7 \\
1.2 & 3.3 & 11.1 & 20.5 & 4.3 & 10.6 & 21.8 \\
1.1 & 2.4 & 7.5 & 14.5 & 3.6 & 8.4 & 17.7 \\
1.0 & 2.6 & 5.7 & 11.5 & 3.1 & 6.9 & 15.0 \\
0.9 & 1.2 & 4.7 & 8.7 & 2.6 & 5.9 & 12.7 \\
0.8 & 1.3 & 5.7 & 7.0 & 1.8 & 5.5 & 10.5 \\
0.7 & 0.9 & 2.4 & 5.4 & 1.6 & 3.7 & 7.9 \\
\enddata
\caption{Detection likelihood ratios with varying $\rm N_{epochs}$ and $\rm S/N_{epoch}$, structured correlation removed, using complete planet templates. This approach to the detection confidence is easy to compute and yields the correct relative detection strengths, but underestimates the absolute confidence compared with modeling-based approaches which account for the width of the planet feature in the likelihood space. Additional epochs significantly improve the detection strength, even when the total signal-to-noise across all epochs is similar.}
\label{detstrengths}
% \tablerefs{}
\end{deluxetable*}

To set rough limits on the equilibrium temperature and photometric contrast required to make a detection, we simulate the thermal emission spectra of planets with temperature structures generated self-consistently for equilibrium temperatures 600 K, 1000 K, and 1400 K. 
We vary photometric contrast by scaling the planet models from 0.7 $\rm R_J$ to 1.3 $\rm R_J$ in steps of 0.1 $\rm R_J$. 
Note that we do not re-compute the planet model for each radius/photometric contrast to account for changes in gravity or temperature--pressure (T--P) profile.
We simulate 10 epochs with $\rm S/N_{epoch}$ of 2500 and 25 and 50 epochs with $\rm S/N_{epoch} = 1500$, for total S/N of 7900, 7500, and 10600, respectively.
The variation in radius corresponds to an $L$-band photometric contrast range of $2.2\times10^{-5}$ to $7.6\times10^{-5}$ for $\rm T_{eq} = 600$ K, $1.6\times10^{-4}$ to $5.4\times10^{-4}$ for $\rm T_{eq} = 1000$ K, and $3.7\times10^{-4}$ to $1.3\times10^{-3}$ for $\rm T_{eq} = 1400$ K. 
This allows the effect of variations in temperature given radius, variations in radius given temperature, and simultaneous variations in both radius and temperature to be distinguished and compared. 
\emph{A priori}, we expect larger and hotter planets should be more easily detected for fixed stellar properties due to the smaller photometric contrast with the host star giving a better signal-to-noise on planet lines.

Previous applications of the 2D multi-epoch cross-correlation technique have found significant non-planetary structure in the observed $\log L(\rm K_p)$ curves, well in excess of the errors estimated from jackknife tests \citep{piskorz88133, piskorzupsand, Piskorz2018, Buzard_2020}. 
These features are believed to arise from a combination of telluric residuals, features in the observed star/planet not represented in the template spectra, and the correlation between the planet template and observed stellar spectrum. 
In some cases \citep[e.g.][]{lockwood, Buzard_2020}, these effects can be comparable in strength to the cross-correlation features caused by the planet, requiring careful identification and modeling to correctly identify the planet signal and complicating estimates of the detection strength.
Proper echelle and cross-disperser settings can also help to minimize these effects by avoiding spectral regions prone to strong telluric residuals or windows where the stellar and planetary spectra are strongly correlated.

As our simulation framework does not include tellurics and uses the same model framework for both cross-correlation templates and simulating observed spectra, the off-peak structure (visible in Figure \ref{brogizuckercomp}) should be dominated by the correlation between the planet template and the stellar spectrum. 
In order to remove structured off-peak correlation, a set of star-only observations containing no planet spectrum was simulated and the resulting $\log \bar{L}(\rm K_p)$ curve subtracted from each of the simulated planet $\log L(\rm K_p)$ curves to give a relative likelihood, $\log RL(\rm K_p)$ (see equation \ref{rleqn}). 
This has the effect of normalizing the likelihood surface by the correlation of the planet model with the observed stellar spectrum. 
The lack of significant structure in Figure \ref{radgridall} between -150 and 0 \kms\ compared with Figure \ref{brogizuckercomp} shows that the subtraction of the star-only simulation almost entirely eliminates the structured off-peak correlation in our simulations. 

Simulations of the stellar spectrum were compared to observations of the combined star/planet spectrum in \citet{Buzard_2020} and successfully reproduced some, but not all, observed non-planetary features in the log-likelihood surface, indicating that the near-perfect correction of structured correlation is not yet achievable in practice. 
A discussion of how this correction may be improved in observations is presented in Section \ref{sec:disc}. 
The inability to achieve good corrections for off-peak correlation in observations complicates the detection of smaller/cooler planets compared with simulations, and the results presented in Figure \ref{radgridall} and Table \ref{detstrengths} likely overestimate the detectability of such planets with current pipelines. 
However, correcting the off-peak correlation enables a more robust comparison between different simulations, and offers insight on the possibilities with future pipeline improvements.

The $\log RL(\rm K_p)$ curves for each $\rm T_{eq}$ and $\rm R_p$ are plotted in Figure \ref{radgridall}, and detection likelihood ratios are listed in Table \ref{detstrengths}. 
Jackknife tests are used to estimate the shaded 1$\sigma$ error region. 
In addition to the total planet spectrum, we perform the cross-correlation with molecule-specific planet templates for H$_2$O, CH$_4$, CO, NH$_3$, H$_2$S, CO$_2$, and PH$_3$, in order to assess the impact of individual species on the detection. 
Only H$_2$O and CH$_4$ are robustly detected, and are plotted in Figure \ref{radgridall} along with the NH$_3$ and complete templates for comparison. 
The relative strength of the H$_2$O and CH$_4$ detections varies with simulated temperature, as expected from the equilibrium chemistry. 
The NH$_3$ template gives a featureless $\log RL(\rm K_p)$ curve in the 1400 K and 1000 K models, but shows weak features in the 600 K case, though not significant in comparison with H$_2$O or CH$_4$. 
The H$_2$S template shows some weak features near the expected planet peak in the 1000 K models, but the detection is similarly not significant compared with H$_2$O or CH$_4$. 
The CO, CO$_2$, and PH$_3$ templates all produce featureless $\log RL(\rm K_p)$ curves after correcting for the stellar contribution in all cases, consistent with the expected lack of $L$-band spectral features. 

In the 600 K simulations, correlating with all molecules in the planet model results in a much stronger detection than would be expected from the individual H$_2$O and CH$_4$ cross-correlation functions. This does not appear to be the result of other molecules contributing substantially to the total detection. Rather, we note that omitting a molecule from the cross-correlation template -- but not the simulated observation -- effectively decreases the signal-to-noise of the observation. The features present in the simulated data but omitted from the template will still correlate with the template spectrum, but at values other than the planetary radial velocity, increasing the off-peak correlation and reducing the relative size of the planet peak. The impact is more pronounced than an increase in the Gaussian noise because the non-random structure produces off-peak structure that will add consistently when epochs are combined, whereas structure arising from random noise will average to zero across epochs.

Detection strength estimates for HRCCS detections have been made in a variety of ways. 
In NIRSPEC observations prior to the 2019 upgrade, \citet{lockwood} reported a 6$\sigma$ detection in 5 epochs by comparing observed and synthetic log-likelihood functions. 
\citet{piskorz88133} reported a strong detection based on 6 epochs by computing the Bayes factor, comparing a Gaussian planet detection with a linear nondetection in the log-likelihood space. 
\citet{piskorzupsand} reported a 3.7$\sigma$ detection in 7 epochs and \citet{Piskorz2018} reported a 3.8$\sigma$ detection in 6 epochs using a similar approach. 
\citet{Buzard_2020} reported a 6.5$\sigma$ detection in 7 epochs, two of which were taken with the upgraded NIRSPEC, by fitting the full observed log-likelihood function with simulations.

Using the Bayes factor technique described in \citet{Piskorz2018} and the complete all-molecule planet template results in $>5\sigma$ detections for all 50-epoch simulations and all 25-epoch simulations except for two cases with $\rm T_{eq} = 600$ K, $\rm R \leq 1.1 R_J$. 
This is consistent with the higher resolution, increased spectral coverage, and increased number of epochs used in the simulations compared with prior NIRSPEC observations, as well as the correction for stellar correlation features clarifying the planet signal. 
The 10-epoch simulations range from $<2\sigma$ to $>5\sigma$, despite roughly the same total S/N as the 25-epoch simulations, indicating that many low-S/N observations yield more reliable detections than a smaller number of epochs with similar total S/N, consistent with the simulations from \citet{Buzard_2020}.

The correction of non-planetary correlation features allows us to compare the Bayes factor approach with a likelihood ratio, which we compute as:
\begin{equation}
    LR = 2\times[\log L(\hat K_p) - \log L(\Bar{K_p})]
\end{equation}
Where $\log L(\hat K_p)$ the the value of the log-likelihood function at the planet peak and $\log L(\Bar{K_p})$ is the median value far from the planet peak. 
The resulting LR values are listed in Table \ref{detstrengths}. 
While these values can be used to compute a $\sigma$ detection confidence using Wilkes' Theorem, doing so fails to account for the width of the planet feature in the log-likelihood space, and underestimates the detection confidence as a result. 
However, the likelihood ratio approach is easier to calculate for a large number of simulations, particularly for very well-detected cases, and still yields the correct {\it relative} detection strengths. 
These factors make the likelihood ratio a more useful measure of how detection confidence will vary with planet properties and observation strategy. 

\subsection{Planet Temperature Simulations}\label{sec:tlim}
% It's putting the figure in the right place for now at least
\begin{figure*}
    \centering
    \noindent\includegraphics[width=39pc]{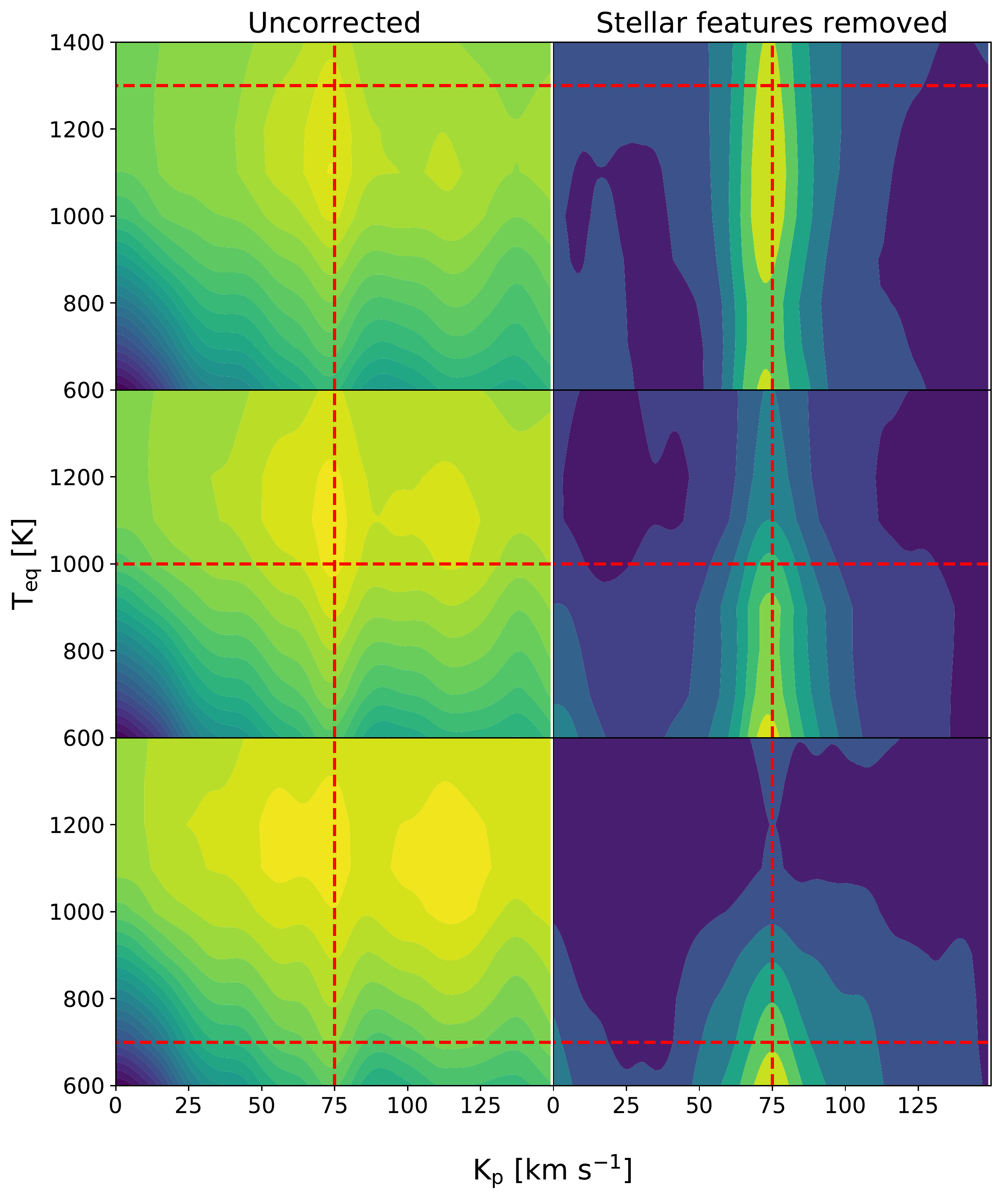}
    \caption{Two-parameter  $\log RL(\rm K_p, T_{eq})$ relative likelihood functions for simulations with equilibrium temperatures of 700 K, 1000 K, and 1300 K. True values are indicated in dashed red. Contours in the left column represent approximately $3\sigma$ changes in the 2D surface while contours in the right column are set at $2\sigma$. The right column subtracts a star-only set of simulated observations to remove structure arising from the stellar spectrum, clarifying the temperature sensitivity. The best-fit $\rm T_{eq}$ is consistently lower than the true value by several hundred K. The preferred value of $\rm K_p$ is nearly independent of temperature.}
    \label{tgridall}
\end{figure*}

We assess the impact of planet equilibrium temperature on the cross-correlation detection by correlating 10 and 25 epochs of simulated observations with $\rm R = 1.0\ R_J$, $\rm S/N_{epoch} = 1500$, and varying planet equilibrium temperatures with models ranging from 600 K to 1400 K in steps of 100 K, covering the ``warm/hot" range of exoplanet temperatures. 
This allows us to compute the two-parameter log-likelihood function $\log L(\rm K_p, T_{eq})$, plotted in the left column of Figure \ref{tgridall} for $\rm T_{eq} = $ 1300 K, 1000 K, and 700 K. 
As before, we also compute the log-likelihood surface from a simulation of the stellar spectrum alone, which is then subtracted in order to remove structure arising from the correlation between stellar and planetary spectra (see equation \ref{rleqn}). 
The resulting log relative likelihood surfaces, $\log RL(\rm K_p, T_{eq})$, are plotted in the right column of Figure \ref{tgridall}.

\begin{figure*}
    \centering
    \noindent\includegraphics[width=40pc]{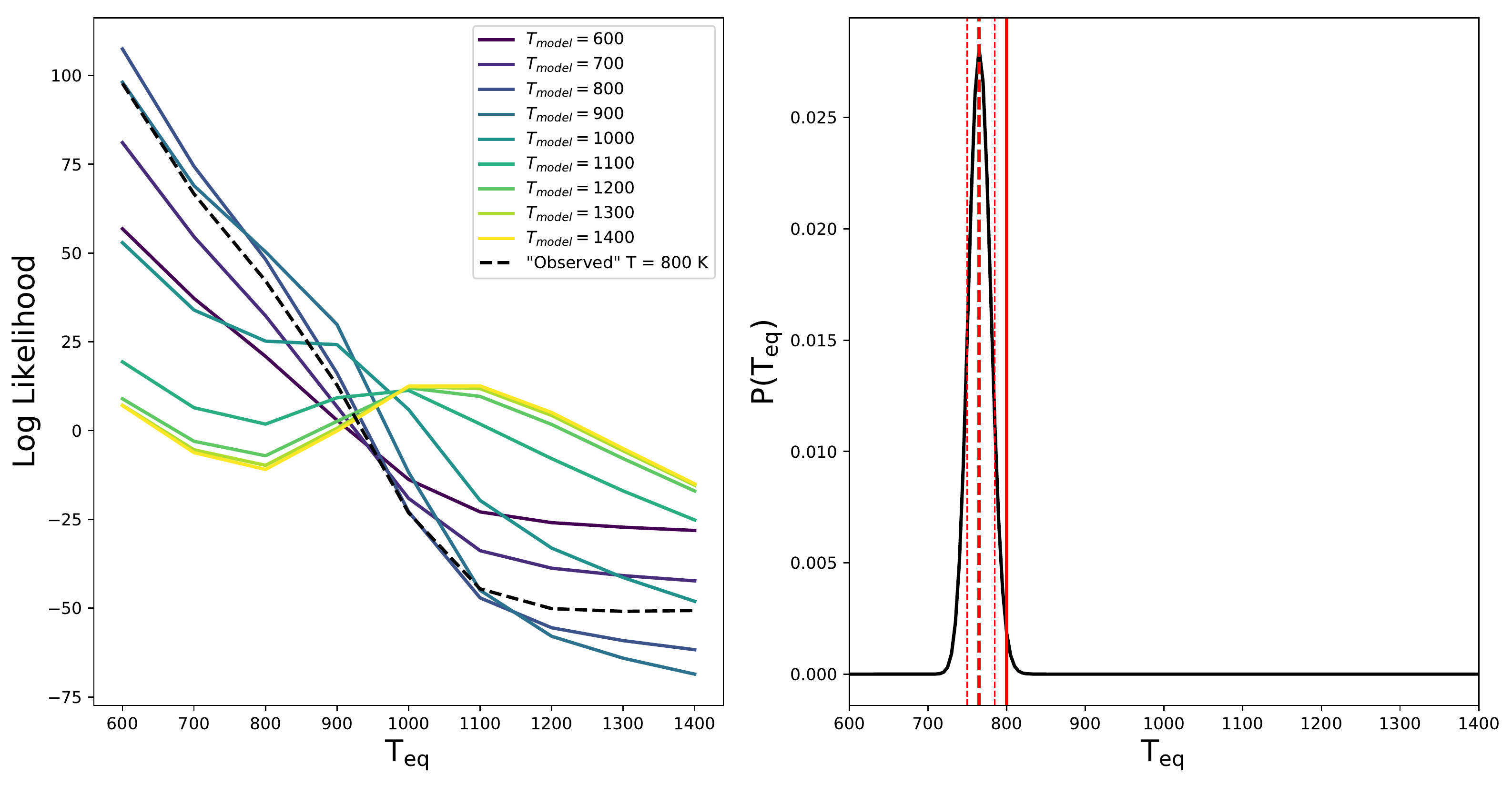}
    \caption{Temperature constraint for 25 epochs of a 1 $\rm R_J$ planet with $\rm T_{eq} = 800$ K.  On the left, the 1D marginalization of the 2D log relative likelihood surface near the best-fit $\rm K_p$ is plotted for the ``observed" spectrum and a grid of models with different temperatures. The right panel plots the probability density $\rm P(T_{eq})$ computed from the deviations between the ``observed" and model curves in the left panel, with the true value indicated in solid red, best fit in dashed red, and 1$\sigma$ errors in dotted red.}
    \label{tfitpeak}
\end{figure*}

We then marginalize the two-parameter $\log RL(\rm K_p, T_{eq})$ surface over the 10 \kms\ region surrounding the best-fit value of $\rm K_p$, roughly corresponding to the resolution of NIRSPEC, to obtain $\log RL(\rm T_{eq})$, plotted in black in the left panel of Figure \ref{tfitpeak}. The size of the marginalization region has minimal impact on the final constraints obtained for simulations in which unwanted star/planet correlation features are effectively removed. We use a 10 \kms\ region for subsequent analysis in order to reflect the uncertainty in the best-fit $\rm K_p$ value while minimizing contamination from residual non-planetary correlation features.
As expected based on the right column of Figure \ref{tgridall}, the $\log RL(\rm T_{eq})$ curve is temperature-dependent but does not show a clear peak at the true value. 
We therefore use a modeling approach to attempt to constrain the temperature, analogous to the technique used by \citet{Buzard_2020} to determine $\rm K_p$ in the presence of significant non-planetary features in the $\log RL(\rm K_p)$ curve. In this case, $\rm K_p$ is already well-constrained (see Figure \ref{tgridall}, right column), and we instead attempt model extraneous features in the $\log RL(\rm T_{eq})$ curve in order to better estimate $\rm T_{eq}$.
A series of models is created with equilibrium temperatures from 600 K to 1400 K in steps of 100 K.  
The two-parameter log-likelihood surface is calculated for each model, and a simulated stellar spectrum is subtracted to remove off-peak correlation features. Finally, we marginalize the models over the same $\rm K_p$ range as was previously used for the ``observed" spectrum.
Both the ``observed" slice we wish to constrain and these model slices are subtracted to have zero mean, accounting for the lack of Gaussian noise in the models, and plotted in the left panel of Figure \ref{tfitpeak}.

While there is not a unique peak at the correct temperature, the $\log RL(\rm T_{eq})$ function still shows significant changes in its shape with model equilibrium temperature. 
Constraining the temperature therefore requires an additional step to compare the observed and expected $\log RL(\rm T_{eq})$. 
We compute the negative log-likelihood of the deviation between the ``observed" curve and each model, under the assumption that the model points are Gaussian-distributed:
\begin{equation}\label{logldev}
    - \log L(T_{eq}) = -\sum_i \frac{(O_i - M(T_{eq})_i)^2}{2\sigma^2(T_{eq})_i}
\end{equation}
Where $O_i$ is the value of the observed log-likelihood function at equilibrium temperature $i$ in the parameter space and $M(\rm T_{eq})_i$ is the value of the model log-likelihood function with equilibrium temperature $\rm T_{eq}$ at temperature $i$. 
$\sigma(\rm T_{eq})_i$ is the error in the associated observed values determined from a jackknife test. To more easily compute best-fit values and confidence intervals, we convert the log-likelihood to a probability distribution $\rm P(\rm T_{eq})$ by exponentiating and normalizing the integral. 
This can be written as:
\begin{equation}\label{loglpdf}
    P(T_{eq}) = C \prod_i \exp\left(\frac{(O_i - M(T_{eq})_i)^2}{2\sigma^2(T_{eq})_i}\right)
\end{equation}
\begin{equation}
   = C \exp\left(\sum_i\frac{(O_i - M(T_{eq})_i)^2}{2\sigma^2(T_{eq})_i^2}\right)
\end{equation}
Where $C$ is a numerically-determined normalization constant so that $\rm P(T_{eq})$ integrates to unity. 
The value of $\rm T_{eq}$ which maximizes $\rm P(T_{eq})$ is the best-fit temperature of the ``observed" planet and the 1$\rm \sigma$ errors on $\rm T_{eq}$ are computed as the interval enclosing an area of 0.68 around the best-fit value. 
These are reported in Table \ref{teffcon}.

Meaningful constraints are obtained for $\rm T_{eq} \leq 1200$ K in both the 10 and 25 epoch cases. 
At higher temperatures, the fitting becomes unreliable, with the 1300 K and 1400 K cases effectively indistinguishable even in the 25 epoch simulations. 
The measured values are generally smaller than the true values, and an increased number of epochs results in negligible improvements to accuracy. 
Additionally, the reported precision based on jackknife estimates of the error in $\log RL(\rm T_{eq})$ appears to underestimate the true uncertainty. 
Comparing the true and measured values suggests the precision is $\sim$50--100 K, though the 100 K resolution of the model grid limits our ability to make a robust estimate.

At the same time, Figure \ref{tgridall} shows that the correct value of $\rm K_p$ is recovered even with significant mismatches in $\rm T_{eq}$ between observed and template spectra, so strong priors are not required in order to make an initial detection. While neglecting the day/night temperature mismatch will have a negative impact on the detection strength compared with models accounting for the full temperature structure, it does not produce an error in $\rm K_p$.
Finally, we note that the detection strength is maximized when the cross-correlation template is cooler than the observed spectrum by $\sim200$ K.

\begin{deluxetable}{ccc}\centering
\tablewidth{0pt}
\tabletypesize{\scriptsize}
\vspace{0.5cm}
\tablehead{\colhead{``Observed" $\rm T_{eq}$ [K]} & \colhead{Measured, 10 epoch} & \colhead{Measured, 25 epoch}}
\startdata
700  & $660\pm20$  & $640\pm10$ \\
800  & $720\pm30$  & $760\pm20$ \\
900  & $850\pm20$  & $850\pm20$ \\
1000 & $1010\pm20$ & $1020\pm20$ \\
1100 & $1100\pm30$ & $1120\pm20$ \\
1200 & $1160\pm90$ & $1180\pm90$ \\
\enddata
\caption{$\rm T_{eq}$ constraints from 10 and 25 epochs of a 1 $\rm R_J$ planet with $\rm S/N_{epoch} = 1500$.  Accurate constraints are obtained for planets with $\rm T_{eq} \leq 1200$ K, though the jackknife estimates appear to underestimate the errors.}
% \tablecomments{}
\label{teffcon}
% \tablerefs{}
\end{deluxetable}

\subsection{C/O Simulations for Warm Planets}\label{sec:co}
\begin{figure*}
    \centering
    \noindent\includegraphics[width=39pc]{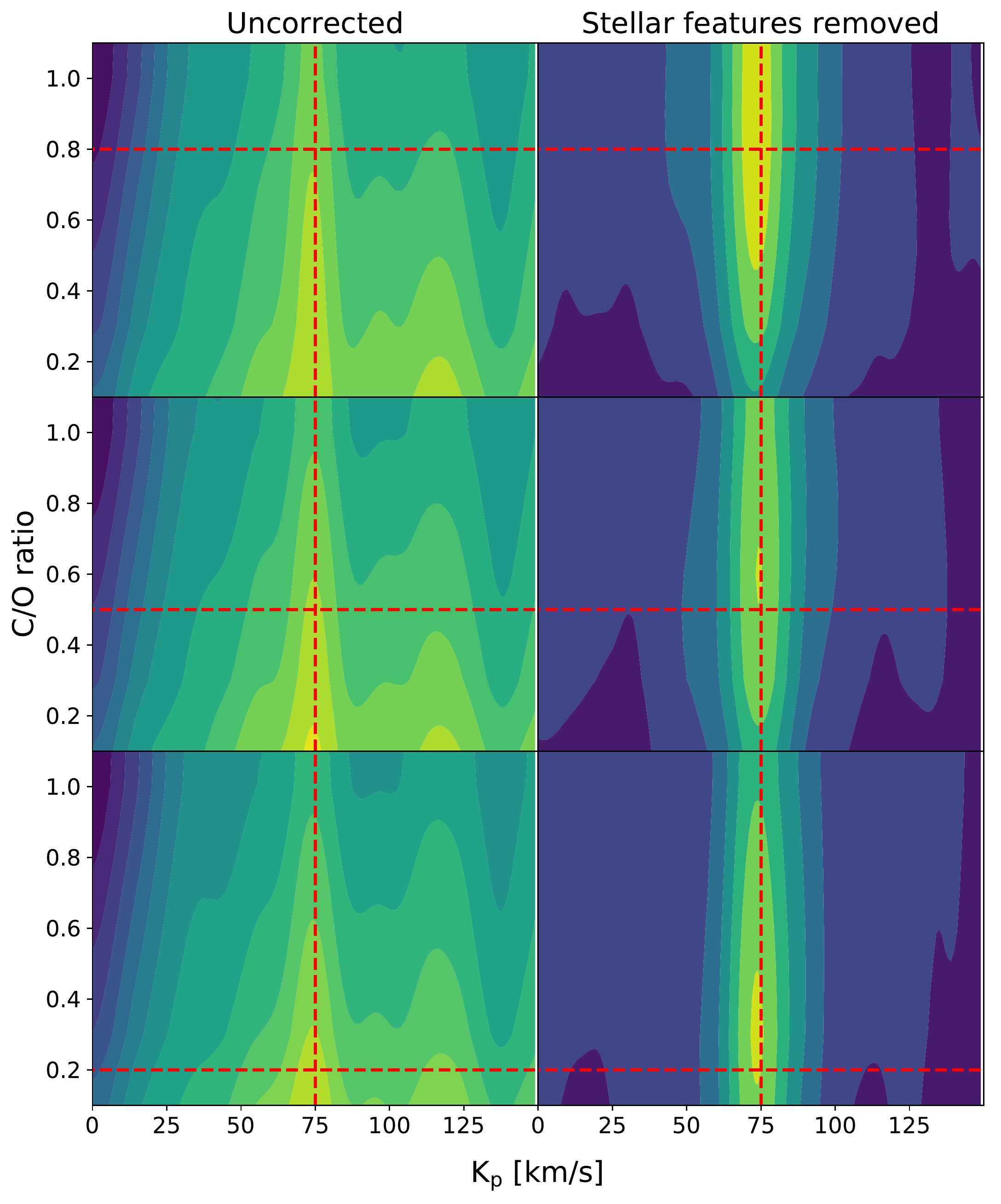}
    \caption{Two-parameter $\log L(\rm K_p,C/O)$ and $\log RL(\rm K_p,C/O)$ log-likelihood functions for simulations with equilibrium temperatures of 900 K and C/O of 0.8, 0.5, and 0.2. True values are indicated in dashed red. Contours in the left column represent approximately $3\sigma$ changes in the 2D surface while contours in the right column are set at $2\sigma$. Stellar correlation features have been subtracted in the right column to produce a relative likelihood. The changes with simulated C/O suggest a sensitivity to the C/O ratio. The recovery of the correct $\rm K_p$ despite significant mismatches between the observed and correlation model C/O ratios indicates strong priors are not required for detection. }
    \label{cogridall}
\end{figure*}

\citet{Piskorz2018} attempted to place a C/O ratio constraint on the hot Jupiter KELT-2Ab using 2D multi-epoch cross-correlation with data from Spitzer and pre-upgrade Keck-NIRSPEC. 
The large errors in this measurement are consistent with Figure \ref{plmodspec} and Figure \ref{radgridall}, which for a hot planet indicate the $L$-band spectrum and resulting cross-correlation function is dominated by H$_2$O features, with no measurable contribution from carbon-bearing species. 
The lack of carbon species lines at high temperature prevents measurement of the C/O ratio. 
However, the simultaneous detection of H$_2$O and CH$_4$ at equilibrium temperatures of approximately 1000 K in Figure \ref{radgridall} suggests a meaningful C/O constraint may be possible for cooler planets at the same wavelengths, assuming chemical equilibrium and known metallicity.

To assess the ability to constrain C/O in cooler planets, we simulate 25  epochs of planets with an equilibrium temperature of 900 K, radius 1 $\rm R_J$, $\rm S/N_{epoch} = 1500$, and a range of C/O ratios, analogous to the simulations in Section \ref{sec:tlim} but varying C/O ratio instead of $\rm T_{eq}$.
We perform a set of simulations with solar metallicity and another with $10\times$ solar metallicity to estimate the impact on the C/O constraint. 
The left column of Figure \ref{cogridall} plots the two-parameter log-likelihoods, $\log L(\rm K_p, C/O)$ for simulations with C/O of 0.8, 0.5, and 0.2 at solar metallicity.  
As with the two-parameter temperature likelihood functions, a star-only simulation was subtracted to reduce structured off-peak correlation, producing the $\log RL(\rm K_p, C/O)$ surfaces plotted in the right column of Figure \ref{cogridall} (see equation \ref{rleqn}). 
The correct value of $\rm K_p$ is recovered regardless of C/O ratio, indicating that a even a large mismatch between template and observed C/O ratio will not result in a non-detection. 
In contrast with the $\log RL(\rm K_p, T_{eq})$ surfaces, the $\log RL(\rm K_p, C/O)$ surfaces appear to be uniquely peaked near the correct value of C/O, after correcting for features arising from the stellar spectrum.

\begin{figure*}
    \centering
    \noindent\includegraphics[width=40pc]{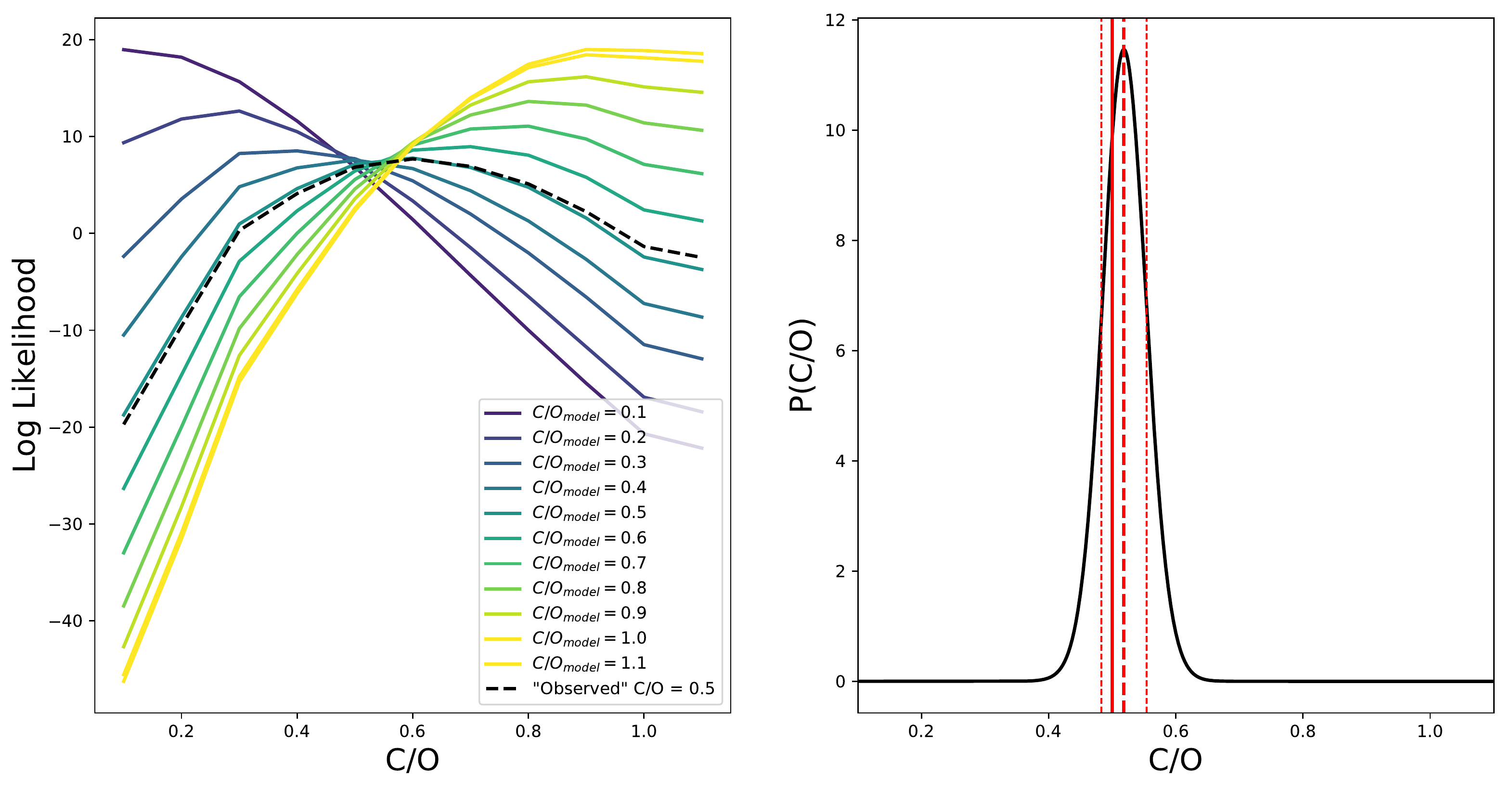}
    \caption{C/O constraint for a simulation with C/O = 0.5 and $\rm T_{eq} = 900$ K. Zero-mean 1D slices of the likelihood surface along the best-fit value of $\rm K_p$ are plotted at left for the target simulation and a series of models. The corresponding probability density function of the ``observed" spectrum having a given C/O, $\rm P(C/O)$, is plotted at right, with the true value in solid red, best-fit in dashed red, and 1$\rm \sigma$ confidence interval in dotted red.}
    \label{copeaks}
\end{figure*}

\begin{deluxetable}{ccc}\centering
\tablewidth{0pt}
\tabletypesize{\scriptsize}
\tablehead{\colhead{``Observed" C/O} & \colhead{Measured C/O}  & \colhead{Measured C/O} \\ & $\rm z = z_\odot$ & $\rm z =  10\times z_\odot$}
\startdata
0.2 & $0.24\pm0.04$ & $0.26\pm0.04$ \\
0.3 & $0.34\pm0.03$ & $0.39\pm0.04$ \\
0.4 & $0.41\pm0.04$ & $0.45\pm0.04$ \\ 
0.5 & $0.52\pm0.04$ & $0.53\pm0.03$ \\
0.6 & $0.60\pm0.04$ & $0.62\pm0.03$ \\
0.7 & $0.82\pm0.06$ & $0.72\pm0.03$ \\
0.8 & $0.86\pm0.06$ & $0.82\pm0.02$ \\
0.9 & $0.96\pm0.08$ & $0.92\pm0.02$ \\
1.0 & $0.92\pm0.08$ & $0.96\pm0.03$ 
\enddata
\caption{Measured C/O values for 25 epochs with $\rm T_{eq} = 900$ K,  $\rm R = 1.0\ R_J$, and $\rm S/N_{epoch} = 1500$, at solar and $10\times$ solar metallicity. Accurate constraints are obtained for C/O $< 0.7$ at solar metallicity and C/O $< 0.9$ at $10\times$ solar metallicity. At high values of C/O, the ``observed" spectra become difficult to distinguish reliably. These results are sufficient to distinguish between sub-stellar, super-stellar, and approximately stellar values of the planetary C/O ratio.}
\label{coconstraints}
% \tablerefs{}
\end{deluxetable}

As was the case with equilibrium temperature, the two-parameter likelihood surface changes with model C/O ratio, indicating the multi-epoch technique is sensitive to observed C/O, and the recovery of the correct $\rm K_p$ at all C/O values suggests strong priors on the C/O ratio are not required to make a detection. 
We use the same approach to characterize the C/O sensitivity as was previously used for equilibrium temperature. 
We begin by marginalizing the 2D $\log RL(\rm K_p, C/O)$ surface over the 10 \kms\ region surrounding the best-fit $\rm K_p$, roughly corresponding to the resolution of NIRSPEC, to obtain $\log RL(C/O)$, which we plot in the left panel of Figure \ref{copeaks}.

While the $\log RL(C/O)$ curves are generally peaked near the correct value, the modeling approach described in the previous section continues to provide more accurate constraints when the models closely replicate observations. 
We therefore correlate a set of noise-free models which we fit to the observed $\log RL(C/O)$ curve following equation \ref{logldev}, which we then convert to $\rm P(C/O)$ following equation \ref{loglpdf}. 
The right panel of Figure \ref{copeaks} plots the resulting $\rm P(C/O)$ function for a simulation with $\rm C/O = 0.5$, analogous to the right panel of Figure \ref{tfitpeak}.

Table \ref{coconstraints} shows that a good constraint on C/O is achieved for input values below C/O = 0.7 from 25 epochs of $\rm S/N_{epoch} = 1500$ and solar metallicity, and below C/O = 0.9 for $10\times$ solar metallicity. 
At higher C/O ratios, the planet spectra become too similar to reliably distinguish between different values for C/O from the $L$-band spectrum alone. 
At lower C/O ratios, the measured values tend to overestimate the true C/O, particularly in the higher-metallicity models. 
Despite this, the measured C/O values are accurate to better than 0.1, which should enable differentiation between substellar, approximately stellar, and superstellar C/O ratios. 
Higher metallicity leads to somewhat better constraints, particularly at high C/O ratios, as the increased heavy element content leads to stronger, better-detected H$_2$O and CH$_4$ lines.
Both accuracy and precision are substantially improved in the 25-epoch simulations compared with 10-epoch simulations of similar total signal-to-noise.

\subsection{Instrument Properties}\label{sec:inst}
\begin{figure}
    \centering
    \noindent\includegraphics[width=19pc]{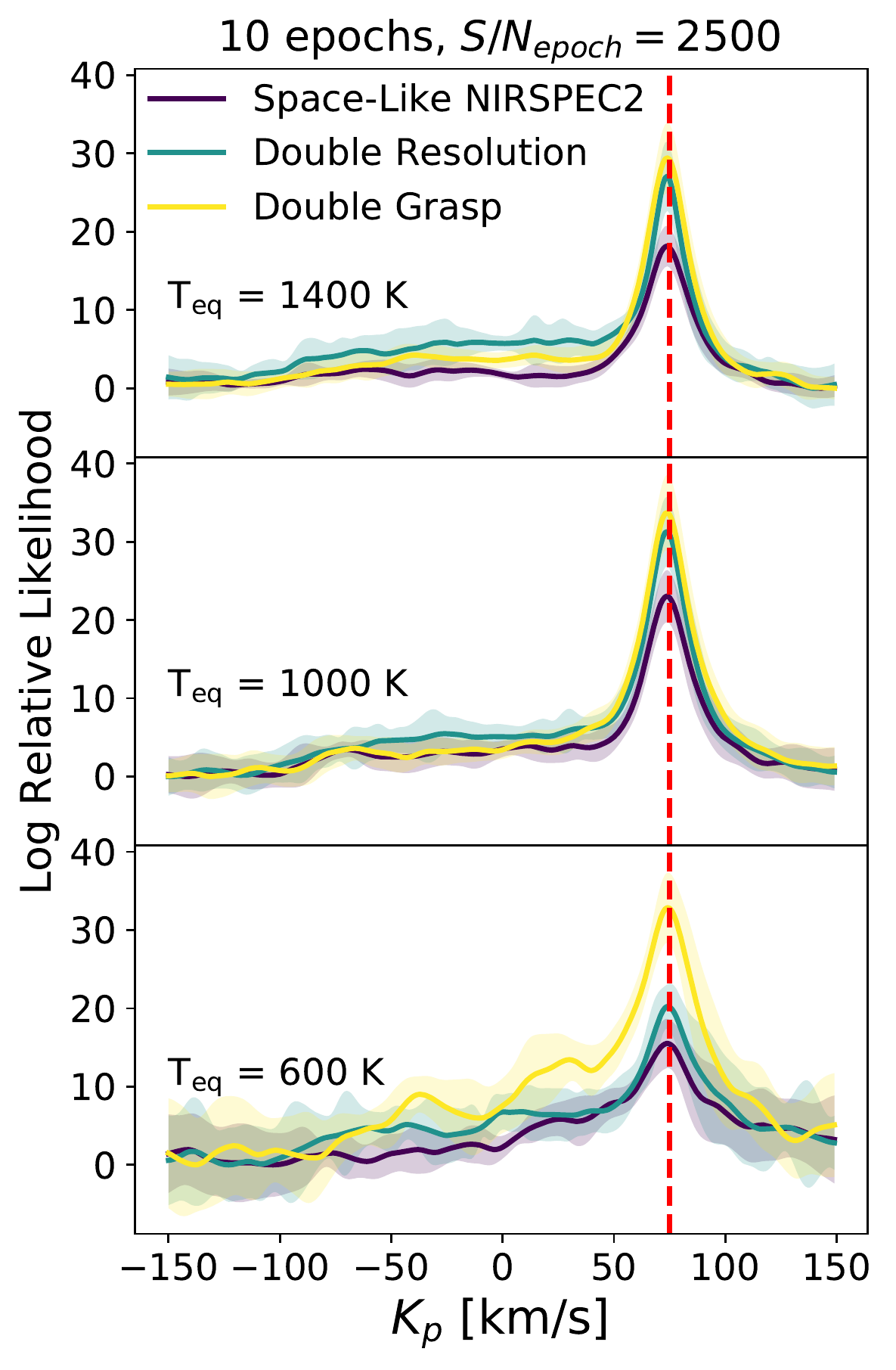}
    \caption{Comparison of the relative likelihood curves for different instrument properties. The top panel plots a simulated planet with $\rm T_{eq} = 1400$ K, the middle $\rm T_{eq} = 1000$ K, and the bottom panel $\rm T_{eq} = 600$ K. Ten epochs are used in each simulation. The NIRSPEC2 case uses the wavelength range plotted in Figure \ref{plmodspec} and R $\sim$ 35000, while the double resolution case considers the same wavelength coverage, but with R $\sim$ 70000. The double grasp case uses R $\sim$ 35000, but with double the wavelength coverage of the NIRSPEC2 case.}
    \label{instplot}
\end{figure}

Our final set of simulations assesses the role of spectral grasp and spectral resolution in both detection and atmospheric characterization. 
In order to compare instrumental properties directly, without considering the impact of the significant absorption features present in the $L$-band, we use ``space-like" simulations in which ixels affected by saturated tellurics are not masked, unlike the simulations presented previously. We consider three cases with ten evenly-spaced epochs each. 
First, we consider a space-based NIRSPEC2 analog. The $\rm S/N_{epoch}$ is 2500, spectral resolution is $R\sim35000$ and the spectral coverage is plotted in Figure \ref{plmodspec}. 
Second, we consider a space-based instrument with the same wavelength range as NIRSPEC2, but with double the resolution ($R\sim70000$). To account for the increased dispersion, $\rm S/N_{epoch}$ is reduced to 1768.
Finally, we consider a space-based instrument with the $R\sim35000$ and $\rm S/N_{epoch} = 2500$, but with the spectral grasp doubled to provide nearly continuous coverage from 2.9--3.8 $\mu$m. 
To maintain a consistent sampling of the line spread function, the latter two cases were run with 4096 pixels instead of 2048 for NIRSPEC2.

For each set of instrument parameters, we simulate a set of planets with $\rm T_{eq} = 1400$ K and radii from 0.7 $\rm R_J$ to 1.3 $\rm R_J$, similar to Section \ref{sec:rsim}, as well as a set with $\rm T_{eq} = 900$ K, $\rm R = 1.0\ R_J$, solar metallicity, and a range of C/O ratios, similar to Section \ref{sec:co}. 
Following the same procedures as in those sections, we present detection strengths  as likelihood ratios in Table \ref{detsinst} and C/O constraints in Table \ref{coconinst}. 

\begin{deluxetable}{cccc}\centering
\tablewidth{0pt}
\tabletypesize{\scriptsize}
\tablehead{\colhead{Planet Radius} & \colhead{NIRSPEC2} & \colhead{Double Resolution} & \colhead{Double Grasp}}
\startdata
1.3 & 24.8 & 37.9 & 41.5 \\
1.2 & 20.9 & 32.9 & 35.9 \\
1.1 & 17.0 & 27.4 & 30.0 \\
1.0 & 14.9 & 21.6 & 25.1 \\
0.9 & 12.1 & 18.1 & 20.2 \\
0.8 & 9.4  & 15.0 & 15.2 \\
0.7 & 7.1  & 11.4 & 12.1 
\enddata
\caption{Relative likelihoods of detection for 10 epochs of a planet with $\rm T_{eq} = 1400$ K, observed with three hypothetical space instruments. The NIRSPEC2 case uses the wavelength range plotted in Figure \ref{plmodspec} and $\rm R\sim35000$ based on NIRSPEC2 observations. Double resolution considers the same grasp as NIRSPEC2, but $\rm R\sim70000$. Double grasp considers $\rm R\sim35000$, but with double the wavelength coverage compared with the NIRSPEC2 case. Both spectral resolution and spectral coverage have a significant impact on detection strength.}
\label{detsinst}
% \tablerefs{}
\end{deluxetable}

Examining first the impact on detection strength, we find that both grasp and resolution lead to a significant improvement in detection. 
This can be seen in Figure \ref{instplot}, which plots the relative likelihood curves for each set of instrumental factors for 10 simulated epochs with $\rm T_{eq} = 1400$ K, $\rm T_{eq} = 1000$ K, and $ \rm T_{eq} = 600$ K.
Comparing with the $\rm T_{eq} = 1400$ K, 10 epoch case in Table \ref{detstrengths}, all detections listed in Table \ref{detsinst} are significantly stronger than the corresponding entries in Table \ref{detstrengths}, indicating that telluric losses are significantly degrading ground-based observations. 
Both increased spectral resolution and increased wavelength coverage offer additional improvements in detection strength, with the double resolution case producing a $\sim60$ percent improvement in the likelihood ratio compared with the ``space-like" NIRSPEC2 and the double grasp case yielding a $\sim70$ percent improvement.
The improvement in detection strength appears to be somewhat dependent on planet temperature. While for the $\rm T_{eq} = 1400$ K and $ \rm T_{eq} = 1000$ K planet models both increased resolution and increased grasp produce similar improvements in detection strength, the $\rm T_{eq} = 600$ K models are much better detected in the increased grasp case.

\begin{deluxetable}{cccc}\centering
\tablewidth{0pt}
\tabletypesize{\scriptsize}
\tablehead{\colhead{``Observed" C/O} & \colhead{NIRSPEC2} & \colhead{Double Resolution} & \colhead{Double Grasp}}
\startdata
0.2 & $0.19\pm0.01$ & $0.19\pm0.02$ & $0.23\pm0.01$ \\
0.3 & $0.34\pm0.02$ & $0.35\pm0.02$ & $0.34\pm0.01$ \\
0.4 & $0.40\pm0.02$ & $0.36\pm0.02$ & $0.40\pm0.02$ \\
0.5 & $0.53\pm0.02$ & $0.52\pm0.02$ & $0.52\pm0.01$ \\
0.6 & $0.64\pm0.03$ & $0.61\pm0.03$ & $0.60\pm0.02$ \\
0.7 & $0.68\pm0.03$ & $0.69\pm0.03$ & $0.70\pm0.02$ \\
0.8 & $0.84\pm0.03$ & $0.87\pm0.04$ & $0.83\pm0.02$ \\
0.9 & $0.98\pm0.06$ & $0.92\pm0.05$ & $0.92\pm0.02$ \\
1.0 & $1.06\pm0.04$ & - & $0.99\pm0.04$
\enddata
\caption{C/O constraints from 10 simulated epochs with varying space-based instrument properties and no tellurics. The NIRSPEC2 case uses the grasp and resolution of previous NIRSPEC2 observations. Double resolution considers the same grasp as NIRSPEC2, but double the spectral resolution. Double grasp considers the same spectral resolution as NIRSPEC2, but with nearly complete wavelength coverage from 2.9--3.8 $\mu$m. Increased wavelength coverage offers better performance improvements, particularly at high C/O.}
\label{coconinst}
% \tablerefs{}
\end{deluxetable}

Table \ref{coconinst} shows the impact of instrument parameters on the C/O measurement. 
While increased spectral range appears to offer some improvements in accuracy, particularly at high C/O, increased resolution does not have a significant impact on the measurements obtained.
We can also estimate the impact of telluric features on the achievable C/O constraint by comparing the space-like NIRSPEC2 case in Table \ref{coconinst} with the values in Table \ref{coconstraints}. 
Both accuracy and precision are substantially worse when saturated telluric absorption features are removed, particularly at higher C/O, consistent with the large loss in effective spectral grasp due to telluric features.

\section{Discussion}\label{sec:disc}
\subsection{Observation Strategy}
Simulations of identical systems with varying number of epochs allow us to directly compare the results of different observation strategies. Table \ref{detstrengths} lists detection strengths for simulations with 10 epochs at $\rm S/N_{epoch} = 2500$ as well as simulations of 25 epochs with $\rm S/N_{epoch} = 1500$. 
The total signal-to-noise ratio in these simulations combining all epochs is $\sim$7900 and $\sim$7500 respectively. 
Despite the slightly better total signal-to-noise, the 10-epoch case gives a substantially worse detection strength in all cases compared with the otherwise-equivalent 25-epoch simulations, corresponding to a factor of $\sim 2$ reduction in photometric contrast. 
Comparing the 25-epoch simulations with 50-epoch simulations, which have 40 percent greater total signal-to-noise, the 50 epoch case performs substantially better that would be expected from the increase in signal-to-noise, again corresponding to a factor of $\sim2$ in photometric contrast. 
This suggests that confidence in the multi-epoch detection is much more dependent on the number of epochs combined than on either the per-epoch or total signal-to-noise. 
This is consistent with findings from \citet{Buzard_2020} that spreading the same total signal-to-noise over an increasing number of epochs results in a stronger detection.

The improvement with increasing number of epochs suggests the optimal observing strategy for multi-epoch cross-correlation is to take a large number of relatively low signal-to-noise observations, similar to stellar radial velocity surveys. 
While stellar optical spectra can be corrected to the required precision for RV measurements by dividing by the spectrum of an A0 telluric standard star, the extremely low planet flux measured in multi-epoch cross-correlation would require a prohibitive amount of time spent telluric standards in order to prevent the uncertainty in the standard measurement from dominating over the planet signal. 

Instead of telluric standards, multi-epoch cross-correlation observations since \citet{lockwood} have made use of line-by-line atmospheric models such as RFM \citep{DUDHIA2017243} or MOLECFIT \citep{Kausch2014} to fit and divide out a model atmosphere. Beginning with \citet{piskorz88133}, this is followed by a principal component analysis (PCA) to correct errors in the model line profile function, changes in molecular abundances over the observation, and instrument flexure \citep{piskorz88133, Piskorz2018}. 
The guided PCA requires a series of observations at varying airmass in order to identify residuals associated with tellurics, as these features should vary consistently with the airmass. Based on prior NIRSPEC observations, a time series covering at least $\sim30$ minutes per target is necessary for guided PCA to be effective. 
This in turn requires the use of the few-epoch, high per-epoch signal-to-noise observing strategy simulated in the 10 epoch case when observing with large telescopes, with $\rm S/N_{epoch} \sim 2500$. 
The results of the simulations presented here suggest that the development of alternative telluric correction procedures which do not impose a minimum integration time should be a high priority for future multi-epoch cross-correlation pipelines. 
We also note that more sensitive instruments may shorten the time series required for effective PCA, but will also further reduce the integration time required to achieve the desired $\rm S/N_{epoch}$ for the many-epoch observing strategy.

\subsection{Photometric Contrast Limits}\label{sec:rlim}
Table \ref{detstrengths} allows us to compare the impact of photometric contrast changes due to differences in equilibrium temperature with changes due to planet radius. 
Comparing otherwise-identical simulations with $\rm T_{eq}$ of 600 K and 1400 K, we find the likelihood ratio is only reduced by $\sim20$ percent in the 600 K case, despite a 94 percent reduction in $L$-band photometric contrast compared with the 1400 K case. 
At the same time, a reduction in photometric contrast of $\sim20$ percent due to planet radius results in a $\sim20$ percent reduction in likelihood ratio at fixed equilibrium temperature, indicating that photometric contrast has a significant impact on detection at fixed equilibrium temperature.
We note that these numbers assume cloud-free models for both temperatures, which is not necessarily appropriate for the 600 K case. The impact of clouds and hazes is discussed in more detail below. 
Additionally, we do not account for changes in the planet spectrum with varying radius. Changes in surface gravity and temperature--pressure profiles in smaller planets are likely to introduce changes in spectroscopic contrast in addition to the change in photometric contrast considered in these simulations.

The difference in the relationship between photometric contrast and detection strength from changes in radius compared with changes in equilibrium temperature emphasizes that multi-epoch cross-correlation is much more sensitive to spectroscopic contrast than photometric. 
On a per-photon basis, the $\rm T_{eq} = 600\ K$ planet models are significantly easier to detect via multi-epoch cross-correlation, due to the presence of deep CH$_4$ features at cooler temperatures. 
However, the presence of additional species in the spectra of cooler exoplanets increases the difficulty of accurately matching the correlation template to the observed spectrum, a challenge which is not present in the simulations.
This suggests anticipating the detectability of exoplanet systems with cross-correlation spectroscopy requires an understanding of the target's thermochemical properties, in particular considering the potential presence and impact of clouds/hazes. 
Figures \ref{tgridall} and \ref{cogridall} indicate a detection can be made despite significant differences between the template and observed spectra, provided the template includes the species present in the observed spectrum.
An initial detection with a poorly-matched model can be subsequently revised by varying the template to maximize detection strength.

Section \ref{sec:rlim} found that the 25-epoch simulations detected nearly all simulated planets with $>5\sigma$ confidence using the Bayes' factor approach described in \citet{piskorz88133}. 
This suggests that using a many-epoch, low per-epoch signal-to-noise observing strategy, $L$-band multi-epoch cross-correlation using NIRSPEC is sensitive to planets with $\rm T_{eq} \geq 600 K$ and $\rm R \geq 0.7\ R_J$ around a Sun-like host star. 
Assuming a Sun-like host star, this corresponds to planets within $\sim0.2$ AU approximately Saturn-size and larger, which should enable the detection and characterization of warm Jupiters. 
While we do not simulate the change in orbital period with semi-major axis, this does not affect the detection provided the sampling of the orbital phase is fixed.
We caution that the exact limits are likely to be strongly dependent on the exact wavelength range observed, the planet chemical composition, the presence of clouds/hazes, and how well features arising from the stellar spectrum can be removed from the $\log L(\rm K_p)$ space.

\subsection{Sensitivity to Temperature}\label{sec:teffsens}
Figure \ref{tgridall} indicates that while mismatches between the true temperature and the correlation model negatively impact the detection strength, the best-fit value of $\rm K_p$ is independent of the model equilibrium temperature over an 800 K range. 
This is particularly relevant to planning observations for tidally-locked planets such as hot Jupiters, which are known to have day/night temperature differences of several hundred Kelvin \citep[e.g.][]{Knutson2007, Wong_2016, Komacek_2016}. 
Figure \ref{tgridall} indicates both dayside and nightside observations can be used effectively in the absence of longitudinally-varying clouds/hazes with a relatively minor impact on the detection strength, though the correlation model should use the lowest estimated temperature. Underestimating the planetary equilibrium temperature in the template spectrum consistently results in a better detection.

While the weak dependence on mismatches between the observed and model temperature is useful observationally, it suggests $L$-band observations have a limited ability to constrain thermal properties of the system.
While Figures \ref{tgridall} and \ref{tfitpeak} do show variations with equilibrium temperature, the values in Table \ref{teffcon} show significant systematic errors. 
Additionally, the temperature constraint in the $L$-band arises primarily from the relative strengths of H$_2$O and CH$_4$ features, which will also depend on metallicity and C/O ratio. 
Multi-dimensional model grids to assess these degeneracies are computationally impractical at present, and will require improvements in the cross-correlation routine.
The $L$-band temperature sensitivity is therefore unlikely to be useful in practice beyond providing rough constraints on the day/night temperature contrast for $\rm T_{eq} \leq 1200$ K, when CH$_4$ features are present in the spectrum. 
However, other NIR spectral features with stronger temperature dependencies, such as CO bandheads in the $K$ and $M$-bands, may offer much better constraints on 3D thermal properties, as was demonstrated in \citet{beltz2020}.
With a large number of epochs, variations in the best-fit temperature between epochs at different orbital phase could enable thermal mapping of non-transiting planets, similar to the phase-curve mapping technique in \citet{Knutson2007}. 
Future simulations will explore the wavelength-dependence of thermal constraints from multi-epoch cross-correlation and the impact of phase-dependent observed planet temperatures. We expect the temperature-dependence of the planetary spectrum will lead to constraints on planet properties varying strongly with equilibrium temperature, even for observations at fixed wavelength.

\subsection{Sensitivity to C/O}
Figure \ref{cogridall} indicates that while the best-fit $\rm K_p$ is unaffected by a mismatched C/O ratio, multi-epoch cross-correlation successfully recovers the true value for $\rm T_{eq} = 900$ K and C/O $\leq 0.8$, as shown in Figure \ref{copeaks} and Table \ref{coconstraints}. 
Higher metallicity offers marginally better performance at high C/O. The contrast with the weak C/O constraint obtained in \citet{Piskorz2018} is due to the presence of CH$_4$ in the $L$-band spectrum for $\rm T_{eq} \leq 1000$ K, which allows carbon and oxygen bearing species to be observed simultaneously. 
This again illustrates the dependence of multi-epoch cross-correlation detection and characterization on the wavelength range observed and the target's atmospheric composition. 
While $L$-band observations can provide useful C/O constraints for warm planets, hot Jupiters require observations at different bands which contain carbon species (e.g. CO) features at high temperature.
Useful C/O constraints are not obtained for simulated planets with $\rm T_{eq} \geq 1000$ K due to the lack of CH$_4$ lines at high temperatures, though for cooler planets differences in $\rm T_{eq}$ between the observed and template spectra do not significantly impact the C/O constraint beyond increasing the uncertainty to $\sim0.1$. 

We note that $L$-band C/O constraints require an equilibrium chemistry assumption, as only H$_2$O and CH$_4$ features are observed. 
\citet{Wallack_2019} found that disequilibrium processes in cooler plants have a minor impact on molecular abundances compared with atmospheric metallicity and C/O ratio for planets with equilibrium temperatures around 1000 K, suggesting the equilibrium assumption is reasonable for the warm exoplanet populations. 
In cases where disequilibrium processes are expected to play a significant role, a CO detection is likely to be required, and disequilibrium processes must be included in the model planet spectrum. 
Large-grasp instruments which can detect multiple carbon and oxygen-bearing species simultaneously would allow large-scale disequilibrium processes to be identified through cross-correlation without \emph{a priori} knowledge of the atmosphere. 

The ability to measure C/O from $L$-band observations has the potential to clarify the formation processes for warm Jupiters. 
\citet{Oberg_2011} found the C/O ratio of the gaseous and solid components of protoplanetary disks varies with distance. 
Gas beyond the H$_2$O snow line is carbon-enriched relative to the inner disk, while solids are carbon-depleted. 
If the gas accretion determines the final C/O ratio, planets which form beyond the snow line and migrate inwards should have significantly superstellar C/O \citep[e.g.][]{Oberg_2011, oberg_2016, madhusudan_2017}. 
Other work has suggested solid accretion dominates the final atmospheric C/O \citep[e.g.][]{mordasini_2016, espinoza_2017}, resulting in substellar C/O values which drop further beyond the H$_2$O snow line. 
The relative importance of gas and solid accretion may also vary significantly with planet mass, leading to compositional differences between Jupiter and Neptune-size planets \citep{cridland_2019}. 
The precision of the C/O constraints in Table \ref{coconinst} suggest cross-correlation spectroscopy can distinguish between migration and \emph{in-situ} formation scenarios, as well as between gas and solid-dominated accretion. 
The weak inclination dependence of cross-correlation techniques results in many more targets for which C/O measurements can be made compared with transit spectroscopy, increasing the potential sample size for studies of atmospheric C/O in intermediate-temperature ($\rm T_{eq}\sim900$ K) planets. 

\subsection{Applicability of Simulations to Observations}
These simulations suggest multi-epoch cross-correlation is capable of significantly more than has been demonstrated observationally. 
In part, this is because these simulations are based on the capabilities of NIRSPEC2 \citep{martin2018}, which offers substantially improved wavelength coverage and resolution compared with the pre-upgrade instrument which was the basis for prior multi-epoch detections. 
As additional observations are taken with the upgraded instrument, future observational results should better match the simulations presented here. 
We also briefly note several factors which were not included in the simulations which impact observations. 
Stellar activity and atmospheric clouds/hazes are discussed in detail below.

Figures \ref{tgridall} and \ref{cogridall} suggest minor inaccuracies in the planetary spectral template temperature or C/O ratio will have minimal impact on the ability to make a detection. 
While inaccuracies decrease the strength of the detection, the correct value of $\rm K_p$ is recovered for $\rm T_{eq}$ between 600 K and 1400 K and C/O between 0.1 and 1.1, regardless of the true values.
Furthermore, we demonstrate that correlating with a grid of planet models can provide constraints on planet properties which are not known \emph{a priori}. 
The one case where differences between observed and template spectra causes significant issues with detection is when simulated observations with $\rm T_{eq} < 1000$ K are correlated with $\rm T_{eq} \geq 1000$ models.
This appears to be due to the absence of CH$_4$ features in warmer models. 
Provided the species which dominate the observed spectrum are represented in the cross-correlation template, errors in the template appear to have a minimal and identifiable impact on the detection.
Errors in line shape or position in the template will lead to a broader, lower-amplitude peak in the cross-correlation function, and may lead to shifts in the best-fit $\rm K_p$ \citep{brogi2019}.
We note that errors in linelists represent a significant additional source of modelling uncertainty, particularly for CH$_4$. 
Using an updated linelist, \citet{gandhi_2020} is unable to reproduce the CH$_4$ detection reported in \citet{Guilluy_2019} for HD 102195 b, indicating the uncertainties in current linelists are large enough to impact both detection and C/O constraints with cross-correlation techniques. 
Future improvements in high-resolution linelists should reduce the impact of this model uncertainty.

These simulations also lack any residuals from the telluric removal procedure, unlike observations. 
\citet{Buzard_2020} showed that such residuals are not necessary to reproduce major non-planetary features in the cross-correlation space, which arise from correlation between the stellar and planetary spectra.  
While the guided PCA approach should effectively remove all features present in a model telluric spectrum or which vary consistently over the observation, fixed differences between the observed line profiles and positions and the telluric model are likely to lead to residuals in observed spectra which are not included in this simulation framework, but which do contribute to off-peak structure in observations.
Developing simulations which accurately include telluric residuals is an ongoing challenge, as it requires quantifying the deviations between observed and model telluric spectra and how such differences vary with airmass and observing conditions.
The near-perfect correction of off-peak structure achieved in these simulations therefore represents a best-case scenario for telluric removal, and we emphasize the importance of accurate telluric correction for successful multi-epoch cross-correlation.

\subsubsection{Stellar Activity}
Stellar activity and the associated spectrophotometric variations are likely to have a significant impact on the ability to detect and characterize planets with multi-epoch cross correlation. 
The left columns of Figures \ref{tgridall} and \ref{cogridall} show that even a perfectly-modelled static stellar spectrum introduces significant non-random structure in the cross-correlation space. 
This structure can be entirely removed in simulations, as the stellar spectrum is perfectly characterized and the only source of non-random structure included.
In observational applications, intrinsic differences between the observed and model stellar spectra and spectrophotometric variability from stellar activity will lead to non-planetary cross-correlation features that cannot be easily removed by subtracting a set of planet-free simulated observations created from a single stellar spectral model.

The impact of stellar variability can be reduced by taking high-cadence observations.
Minimizing the time between observations also minimizes changes in the stellar spectrum, reducing the resulting variable structure in the cross-correlation function. 
Previous multi-epoch cross-correlation detections have used data taken over three to eight years \citep{piskorz88133, piskorzupsand, Buzard_2020}, covering a large fraction of the 11--year Solar magnetic cycle and longer than the magnetic cycles observed other stars \citep[e.g.][]{donati_2008, morgenthaler_2011}.
Such long baselines are not necessary for warm/hot systems with periods of $<100$ days, but allow significant changes to occur in the stellar spectrum compared with observations taken over shorter periods.

Stellar activity can also be addressed in the analysis pipeline. 
While currently the cross-correlation is performed using a single stellar template for all epochs, using different templates at each epoch could account for changes in the stellar spectrum caused by stellar activity. 
This would require an accurate estimate of the stellar activity at each epoch as well as high-resolution spectral models which incorporate varying activity levels.
Such modeling would also be benefited by high-cadence observations, which may allow a Gaussian processes approach to fit spectral changes due to stellar activity, similar to the approach described in \citet{Rajpaul_2015} to reduce errors in RV observations arising from stellar activity.

\subsubsection{Clouds and Hazes}
All simulations used planetary spectral templates without clouds or hazes, which are likely to be present in observed systems with $\rm T_{eq} < 1000\ K$. 
While clouds are responsible for the flat transmission spectrum in GJ 1214 b \citep{kreidberg_2014}, cross-correlation techniques are sensitive to emission features, and are not necessarily impacted in the same way. 
However, 3D models of hot Jupiter atmospheres find clouds lead to muted high-resolution emission features \citep{harada_2019}, and models of super-Earth atmospheres find clouds may result in blackbody-like spectra with weak line features \citep{morley_2015}. 
These findings suggest the presence of significant cloud cover is likely to have a negative impact on the ability to make detections with high-resolution cross-correlation techniques compared with our simulations, similar to the impact on transiting planets.

Hazy planets, in contrast, may be easier to detect in some cases. 
Models of hazy super-Earths in \citet{morley_2015} found hazes can cause temperature inversions which produce stronger infrared emission features than similar planets without hazes, depending on irradiation, particle size, and haze coverage. 
Some exoplanets with featureless transmission spectra due to hazes may nevertheless be amenable to cross-correlation detection of the emission spectrum in the 1--5 $\mu$m spectral range. Determining the precise impact of clouds and hazes will require additional simulations with a range of cloud/haze properties and compositions. 
These simulations will also need to account for variations in cloud/haze coverage over the atmosphere and the viewing angle of the observer at each epoch, rather than using a single model planet spectrum for all epochs. 
We leave these simulations to future work.

\subsection{Instrumental Factors}
Table \ref{detsinst} and Figure \ref{instplot} indicate improvements to either spectral resolution or spectral grasp lead to significant improvements in detection strength. 
While the increased coverage offers a somewhat greater performance improvement compared with increased resolution for warmer equilibrium temperatures, improved spectral grasp offers substantially better performance than increased spectral resolution for the $\rm T_{eq} = 600$ K case.
We believe the relative importance of resolution and grasp is likely to depend on both the observed wavelength and properties of the target planet, in particular the width of the observed planetary emission features. We leave a more detailed exploration of instrumental factors in detection strength considering a broader portion of the near-infrared spectrum to future work.
Table \ref{coconinst} shows the atmospheric characterization from $L$-band features is much less dependent on instrument properties. 
Improving the C/O constraint is likely to require the ability to simultaneously detect additional species beyond H$_2$O and CH$_4$, necessitating a significant increase in spectral grasp rather than additional improvements in spectral resolution.

Several instruments have been proposed or are being developed to offer single-shot 1--5 $\mu$m coverage with spectral resolution greater than Keck-NIRSPEC, including GMTNIRS, CRIRES+, and IGNIS. 
In addition to improving the detectability of planets, large simultaneous wavelength coverage is likely to offer significant improvements in the ability to characterize planets by enabling the simultaneous detection of additional molecules. 
The large spectral ranges of these instruments will be especially valuable in cases of significant non-equilibrium chemistry, clouds or hazes, or \emph{a prior} uncertainty in the atmospheric composition.

Finally, we note that the space-like NIRSPEC2 simulations in Table \ref{detsinst} result in a factor of $\sim$5 improvement in the likelihood ratio compared with the analogous 10-epoch simulations in Table \ref{detstrengths} which remove portions of the spectrum affected by telluric absorption. 
A space-based high-resolution spectrograph covering the 1--5 $\mu$m spectral range would offer significantly greater capability to detect and characterize exoplanets through cross-correlation spectroscopy compared with ground-based facilities.

\subsection{Applications to Late-Type Stars}
These simulations used a Sun-like star for the host stellar spectrum.  
For a given planet radius and equilibrium temperature, the relative brightness of the planet increases with decreasing stellar radius and temperature. 
From a purely photometric viewpoint, we would therefore expect to make stronger detections around later-type stars, as well as the ability to detect smaller and cooler planets. 
However, the results from Section \ref{sec:rlim} indicate multi-epoch cross-correlation is much more sensitive to spectroscopic contrast than photometric. 
Increased stellar activity and stronger molecular lines in late-type stars may cause spectroscopic contrast to diverge significantly from photometric contrast, impacting planet detectability in ways that are difficult to predict without dedicated modeling.
A rigorous assessment of cross-correlation techniques around K and M type primary stars is currently limited by the lack of sufficiently accurate high-resolution spectral models for late-type stars.  

The development of improved models for late-type stars would further expand the population accessible to characterization through multi-epoch cross correlation with existing instrumentation. 
Of particular interest is planets falling in the radius valley identified by \citet{Fulton_2017} near $\rm 1.8\ R_\oplus$. 
While our simulations show such planets are not likely to be detectable with Keck-NIRSPEC around Sun-like stars, the increased relative brightness of exoplanets around M-dwarfs may enable detections, provided accurate stellar models are available and the atmospheric spectrum offers sufficient spectroscopic contrast. 
Cooler planets may pose additional challenges for telluric correction as the emission spectrum of the planet becomes more similar to the terrestrial spectrum.
The ability to probe the atmospheric composition of these planets with multi-epoch cross-correlation could offer a new avenue to investigate the evolutionary processes affecting intermediate-mass, highly irradiated planets near the transition between rocky and gaseous compositions.

\section{Conclusions}\label{sec:conc}
Our simulations indicate that the multi-epoch cross-correlation approach can be used to detect and characterize a much larger population of non-transiting planets than has been previously studied. In particular, we find:

\begin{enumerate}
    \item Planets with $\rm R_{pl} \geq 0.7\ R_J$ and $\rm T_{eq} \geq 600$ K should be detectable around Sun-like stars with existing instruments in the $L$-band, provided the stellar contribution to the cross-correlation function can be effectively removed. Cooler planets are much more detectable than suggested by the photometric contrast. Detections are significantly improved through additional epochs, even if total signal-to-noise is held constant.
    \item $L$-band cross-correlation spectroscopy is weakly dependent on temperature. While day/night temperature differences have some negative impact on the detection strength using a single-temperature model, such differences are unlikely to prevent detection in the absence of clouds. Precise measurements of thermal properties will require observations at other bands, though $L$-band observations may be able to estimate day/night temperature contrast with future pipeline improvements.
    \item $L$-band observations can provide good constraints on the atmospheric C/O ratio for planets with $\rm T_{eq} \approx 900$ K. Such constraints require the simultaneous detection of carbon and oxygen bearing species. The lack of $L$-band CH$_4$ features in hot Jupiters requires observations at other wavelengths to make a C/O measurement for $\rm T_{eq} \geq 1000\ K$. 
    \item Improvements in both spectral resolution and spectral grasp compared with NIRSPEC2 result in improved planet detection and atmospheric constraints. Future improvements in instrumentation will further expand the population of planets which can be detected and characterized with multi-epoch cross-correlation. The simultaneous detection of additional species should enable more robust constraints on atmospheric properties, particularly clouds/hazes and non-equilibrium chemistry.
\end{enumerate}

In these simulations, we used a portion of the $L$-band for which prior NIRSPEC2 observations were available.  
This allows the actual performance of the detector and losses to saturated tellurics to be replicated in simulation, increasing our confidence that the results presented here are obtainable in practice.  
As additional NIRSPEC2 observations are taken at different portions of the near-infrared, it will be possible to perform these simulations at other wavelengths, allowing us to explore how planet property constraints from multi-epoch cross-correlation depend on wavelength and anticipate the capabilities of future high-grasp instruments.

\acknowledgments{
We thank the anonymous reviewer for their helpful suggestions to improve this paper. The simulations presented herein made use of data obtained at the W. M. Keck Observatory, which is operated as a scientific partnership among the California Institute of Technology, the University of California and the National Aeronautics and Space Administration. The Observatory was made possible by the generous financial support of the W. M. Keck Foundation. 

The authors wish to recognize and acknowledge the very significant cultural role and reverence that the summit of Mauna Kea has always had within the indigenous Hawaiian community.  We are most fortunate to have the opportunity to conduct observations from this mountain.

This work was partially supported by funding from the NASA Exoplanet Research Program (grant NNX16AI14G, G.A. Blake P.I.). L.F. acknowledges the support of the Lynne Booth and Kent Kresa SURF fellowship. S.P. acknowledges funding from the Technologies for Exo-Planetary Science (TEPS) CREATE program.  B.B. acknowledges financial support from the Natural Sciences and Engineering Research Council (NSERC) of Canada and the Fond de Recherche Québécois-Nature et Technologie (FRQNT; Québec).
}
\facilities{Keck:II (NIRSPEC)}
\software{astropy \citep{astropy2013, astropy2018}}

{\footnotesize
\bibliography{sample}}
\bibliographystyle{aasjournal}

\end{document}